\documentclass[aps,amsmath,amssymb,twocolumn,showpacs]{revtex4}
\usepackage{graphicx,color}
\graphicspath{{./}{./figures/}}
\usepackage{dcolumn}
\usepackage{bm}
\usepackage{subfigure}
\usepackage[english]{babel}
\newcommand{\be}{\begin{equation}}
\newcommand{\ee}{\end{equation}}
\newcommand{\bea}{\begin{eqnarray}}
\newcommand{\eea}{\end{eqnarray}}

\newcommand{\rmd}{\mathrm{d}}

\newcommand{\hG}{\hat{G}}

\def\bea#1\eea{\begin{align}#1\end{align}}

\begin{document}

\title{Growth over time-correlated disorder: a spectral approach to Mean-field}
\author{Thomas Gueudre} \affiliation{Politecnico di Torino} 
\date{\today\ -- \jobname}


\begin{abstract}
We generalize a model of growth over a disordered environment, to a large class of It\=o processes. In particular, we study how the microscopic properties of the noise influence the macroscopic growth rate. The present model can account for growth processes in large dimensions, and provides a bed to understand better the trade-off between exploration and exploitation. An additional mapping to the Schr\"ordinger equation readily provides a set of disorders for which this model can be solved exactly. This mean-field approach exhibits interesting features, such as a freezing transition and an optimal point of growth, that can be studied in details, and gives yet another explanation for the occurrence of the \textit{Zipf law} in complex, well-connected systems. 
\end{abstract}
\maketitle

Growth is amongst the most evident properties of complex systems, and pervades studies ranging from econometry \cite{grossman1993innovation,cavalcanti2015commodity} and state policies \cite{bouchaud2015growth,fatas2013policy} to cell biology \cite{monod1949growth} and genetics \cite{brunet2012genealogies}. An outcome either to be sought after (the holy grail of economy) or to be impeded (in tumors and epidemics), it nonetheless remains notoriously difficult to measure and predict \cite{taleb2007black}. This is partly because, in numerous cases, growth is dynamically shaped by environmental cues, for example when foraging resources \cite{chupeau2016random}, balancing a portfolio of assets \cite{bouchaud2003theory}, choosing the next step in a chess game \cite{lai2015giraffe} etc.  How do populations adapt to their ever changing environment? How do they tune exploration strategies to cope with uncertainty? Those questions have received constant attention (see \cite{cohen2007should} and references therein). The difficulties are particularly stringent in complex systems, where growth is at heart an emergent quantity, the macroscopic result of microscopic entities. Choosing the level of description of the system, and understanding to which extent its conclusions are valid  -or so to say universal-, is a standard conundrum \cite{vespignani2012modelling}; macroscopic, phenomenological models often disregard crucial factors, such as the uneven repartition of growth amongst the population. Yet overly detailed models are difficult to manipulate -let alone to solve-, brittle and little informative. 

\begin{figure}[hbt!]
\includegraphics[width=.45\textwidth]{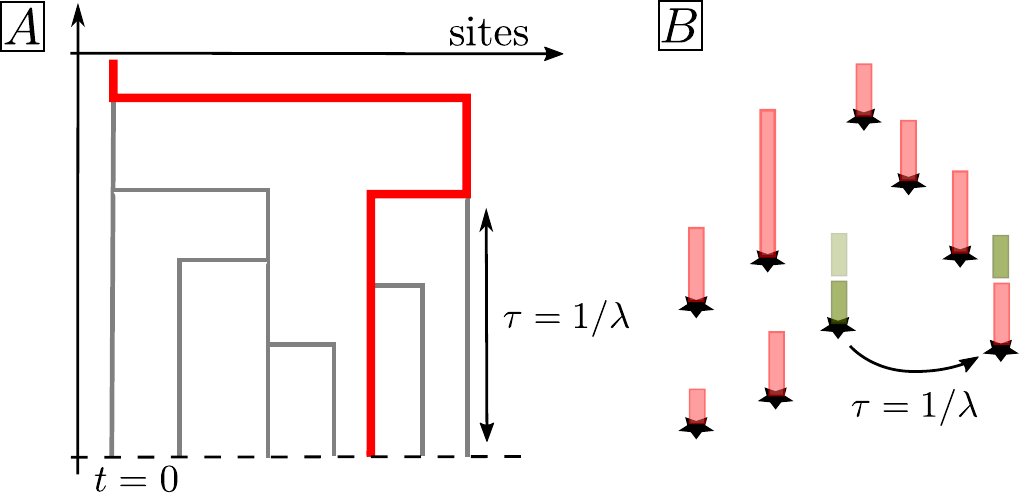} 
   \caption{\textbf{(Sketch) Illustration of the mean-field model of exploration/exploitation}. (Left) Polymers on tree, a problem developped in \cite{DerridaSpohn1988} for tackling the spin glass problem. (Right) Populations on sites randomly exchange members with other sites so that the exchange process conserves the total number. This amounts to implement a discrete Laplacian of infinite range and strength $\lambda$.}
   \label{sketch_AB}
\end{figure}

In the present work, we study in analytical details the influence of the environment randomness over the growth rate. To that purpose, we will consider a \textit{branching random walk}: a population $Z_i(t)$ lives on the nodes $\lbrace i \rbrace_{i \in N}$ of a graph and grows under multiplicative noise $\eta(t)$ (see Fig.\ref{sketch_AB}). We will make the only assumption that $\eta$ is an It\=o process, equipped with a stationary distribution $Q(\eta)$. 
The exploration is implemented by adding a branching mechanism: for each site, at each time step $dt$, the population $Z_i(t)$ may split into two (or more) parts and spread evenly on sites chosen at random with a probability $\lambda \, dt$. The diffusion is of infinite range and this models therefore belongs to the mean-field class. This approach is adapted to describe very connected worlds (such as on complete graphs). It is also relevant in reinforcement learning, where the most common heuristics to solve the exploration/exploitation dilemna, the $\epsilon$-greedy strategy \cite{sutton1998reinforcement,cohen2007should,pratt2006tunable}, is to pick a new strategy at random with probability $\epsilon$. As for now, the question of the optimal $\epsilon$ remains fully open \cite{tokic2010adaptive}. 

We extend a study first started in \cite{gueudre2014explore}, where the above model was solved for a specific It\=o process. A crucial ingredient in its analytical treatment comes from the community of spin glasses \cite{DerridaSpohn1988,brunet2011branching}. In a series of works, Derrida and coworkers unravelled a deep connection of the \textit{branching Brownian motion} to the travelling wave solution of the so-called Fisher-Kolmogorov-Petrovsky-Piscounov (FKPP) equation, a non-linear partial differential equation, most often seen in diffusion-reaction context \cite{grindrod1996theory}. This mapping has since become a standard tool to tackle the branching Brownian motion \cite{arguin2013extremal}. 

The aim of this paper is computational, as we present a large class of growth models for which analytical expression of the growth can be otained. Asides from Gittins indices \cite{Exp3}, often impossible to compute exactly, the panel of solvable growth models exhibiting the exploration/exploitation dilemna is still scarse. This fact has to be compared with the jungle of existing stochastic processes \cite{gardiner1985handbook,pavliotis2014stochastic}, whose enormous development has been triggered by dire modelling needs. Such demand for growth models is apparant in the vast literature on calibrating them onto real datas \cite{gomme1993money,cooley1995economic,janssen1995calibration,caraglio2016export}. Finally, we mention that these results also have interesting implications in other applications of branching random walk that we have not touched upon, such as log-correlated potentials \cite{fyodorov2008freezing,fyodorov2009statistical}, Liouville Field Theory \cite{zamolodchikov2007lectures} or random matrices \cite{fyodorov2014freezing}.

The paper is organized as follows: in Section \ref{FKPP_part}, we first derive the generalized FKPP equation for It\={o} processes. In Section \ref{general_sect}, we compute various asymptotic behaviours to bring out the important quantities related to $\eta(t)$. A standard mapping of the Fokker-Planck onto a Schr\"odinger equation also gives immediately a classification of exactly solvable models  \cite{cooper1995supersymmetry}, and in Section \ref{exactly_solvable}, we illustrate those findings on two such models, rederiving the results of \cite{gueudre2014explore} in a more direct way. Finally in Section \ref{discussion}, we discuss the existence and interplay of various features of growth in this class of models, namely the condensation transition, the optimal growth point and the occurrence of a Zipf law. We also quickly comment on the limitations of the present method.

\section{Evolution equation of the growth rate \label{FKPP_part}}

\subsection{Conventions for the disorder}

We will first introduce the details of the disorder, making the assumption that the resources $\eta(x,t)$ obey an It\={o} equation with time independent drift and diffusion:
\begin{align}
d \eta(t) =  D_1(\eta) \,dt + \sqrt{2 D_2(\eta)} \,\,dW_t
\label{def_Ito}  
\end{align}
where $W_t$ is a Brownian process. The probability distribution $P(\eta,t)$ obeys the Fokker-Planck equation (with the It\={o} prescription):
\begin{align}
\frac{\partial P(\eta,t)}{\partial t} &= - \frac{\partial}{\partial \eta} (D_1 (\eta) P(\eta,t)) + \frac{\partial^2}{\partial \eta^2} (D_2(\eta) P(\eta,t)) \nonumber \\
&= \mathcal{L}_0 P(\eta,t)
\label{fokker_planck}
\end{align}
$\mathcal{L}_0$ being the Fokker-Planck operator of the disorder is written. As commonly stated \cite{risken1984fokker}, the $\eta$ dependence of the diffusion $D_2(\eta)$ can be absorbed by a change of variable and we assume it constant in the following. We also assume $\eta$ to have a stationary distribution:
\begin{align}
Q(\eta) &= \mathcal{N}^{-1} e^{- \Phi (\eta)} \text{   with   } \mathcal{N} = \int d\eta e^{-\Phi(\eta)} \nonumber \\
\Phi(\eta) &= \log D_2 - \int^{\eta} \frac{D_1(u)}{D_2} du \nonumber
\end{align}
with natural boundary conditions (or reflecting in case of bounded support). 
We comment on this hypothesis later in Section \ref{discussion}. As $\Phi$ is defined up to a constant, it can be written as $f$:
\begin{align}
\Phi(\eta) &= f(\eta)/D_2 \\
f(\eta) &= - \int^{\eta} D_1 (u) du
\end{align}
with $f$ the potential of the process. 

Finally, a non zero mean $\mu=\int \eta Q(\eta) d\eta$ simply adds a constant contribution to the growth rate. We set such mean to $0$ and focus on the contribution stemming from the fluctuations of the disorder.

We will especially examine the interplay between exploration and time correlations. To quantify those correlations, it is natural to introduce the -normalized- integrated time correlation function \cite{risken1984fokker}:
\begin{align}
T=\int_0 ^{\infty} \frac{K_{\eta}(t)}{K_{\eta}(0)} dt \nonumber \\
K_{\eta}(t) = \langle \eta(t) \eta(0) \rangle_Q \nonumber
\end{align}
where $\langle \cdots \rangle_Q$ denotes in the following the average with respect to $\eta$. $T$ can be expressed in terms of the terms in Eq.\ref{def_Ito} as \cite{jung1985correlation}:
\begin{align}
T=\frac{1}{K_{\eta}(0)}\int_{-\infty} ^{\infty} \frac{dx}{D_2(x) Q(x)} \left(\int _{-\infty} ^x s Q(s) \, ds\right)^2
\label{risken-jung}
\end{align}
In the following, we will start all our stochastic processes at stationarity, so $K_{\eta}(0)$ reduces to the variance $\langle \eta^2 \rangle_Q$ of $Q(\eta)$.

\subsection{The evolution equation of the growth process}

We consider a large number $N$ of sites, each populated by $Z_i(t)$ elements and resources $\eta_i(t)$, $i=1,... ,N$. 
According to the rules presented in the introduction, each $Z_i(t)$ evolves as:
\bea
\label{model_branching}
Z_i(t+ dt) = 
\begin{cases}
Z_i(t) \exp\left[\eta_i(t) dt  \right]& \text{prob.}\quad 1-\lambda\, dt \\
\quad  \\
\frac{1}{2} (Z_i(t) + Z_j(t))&  \text{prob.}\quad \lambda\, dt
\end{cases}
\eea
where $j \neq j $ labels a site chosen at random amongst the rest. There is considerable freedom in choosing the branching process. We stick to the most common Poisson branching, with a fixed rate $\lambda$, but the derivation below can be easily generalized (see \cite{gueudre2014explore} for some examples).

Owing to its wide fluctuations, the magnitude of $Z_i(t)$ can be estimated in two ways: picking one realization of the disorder and considering its almost sure behaviour  (the quenched setting), or averaging $Z_i$ over all possible realizations of the disorder (the annealed setting). Therefore central quantities are the \textit{typical} growth rate $c_q$  and the more common \textit{average} growth rate $c_a$:
\begin{align}
c_q =\frac{1}{t\,N}  \left\langle  \sum_j \log Z_j(t)  \right\rangle \nonumber \\
c_a = \frac{1}{t \,N} \sum_j \log \left\langle Z_j(t) \right\rangle \nonumber
\end{align}
The first quantity, although harder to calculate, is more representative of the typical, most likely, growth, and we focus on it, following the approach of Derrida and coworkers \cite{DerridaSpohn1988,CookDerrida1990} and \cite{gueudre2014explore} and defining the generating functions, for any $i$:
\bea
\nonumber
G_t(x,\eta) :=& \left\langle \exp\left[-e^{-x}Z_i(t)\right]\delta\left[\eta_i(t)-\eta\right] \right\rangle \\
\label{def_full_G}
\hG_t(x) :=& \int_{-\infty}^\infty \rmd \eta\, G_t(x,\eta) = \left\langle \exp\left[-e^{-x}Z_i(t)\right]\right\rangle \\
G_{t=0}(x,\eta) &= \exp(-e^{-x})\,Q(\eta)
\label{init_cond}
\eea
We assume the disorder initialized at stationarity and $Z_i(t=0)=1$.
Due to the temporal persistance of the disorder $\eta$, we need to keep track of its value through $G_t(x,\eta)$. For long times, the behaviour of $\hG$ is akin to that of a wave front, traveling with constant velocity in the $x$ direction, and reaching $1$ exponentially for large $x$:
\bea
\label{ansatz_derrida}
\hG_t(x) \approx 1- e^{-\gamma(x-c t)} \quad \text{for} \quad x \to \infty.
\eea
Moreover, for $x \rightarrow +\infty$, the generating function goes rather flatly to $0$ and so:
\begin{align}
G_t(+ \infty,\eta) \sim \langle \delta [\eta_i(t) - \eta] \rangle = Q(\eta) 
\label{asymp_ansatz}
\end{align} 
This suggests to look at the following Ansatz for $G_t(x,\eta)$, where dependence in $x$ and $\eta$ are factorized \cite{gueudre2014explore}:
\bea
G_t(x,\eta) \simeq Q(\eta) - R(\eta)e^{-\gamma(x-ct)} \label{ansatz_bouchaud}
\eea
under the constraints:
\bea
\int_{\eta} Q(\eta) d\eta = \int_{\eta} R(\eta) d\eta = 1 \label{norm_const}
\eea
Combining the definition Eq.\ref{def_full_G} with the evolutation equation Eq.\ref{model_branching} and averaging over the disorder, one obtains the following evolution equation for $G_t$:
\bea
\nonumber
& G_{t+\rmd t}(x,\eta) = (1-\lambda\, dt)\, \times \nonumber \\
&\left\langle \exp\left[-e^{-x+\eta_i(t)dt}Z_i(t)\right]\delta\left[\mathcal{G} \, \eta_i(t) -\eta\right] \right\rangle \nonumber \\
\nonumber
& + \lambda \, dt\left\langle \exp\left[-e^{-x-\log(2)}Z_i(t)\right]\delta\left[\eta_i(t)-\eta\right] \right\rangle \times \\
&\left\langle \exp\left[-e^{-x-\log(2)}Z_j(t)\right]\right\rangle
\eea
with $\mathcal{G}$ the infinitesimal propagator of $\eta_t$ over an increment of time $dt$. 
Expanding the arguments of $G_t$ for small $\rmd t$:
\begin{align}
\nonumber
\partial_t & G_t(x,\eta) =  \mathcal{L}_0 G - \eta \partial_x G  \\
\label{hatG_eq}
& + \lambda \left( G_t(x - \log(2),\eta) \hG_t(x-\log(2)) -  G_t(x,\eta) \right)
\end{align}
with $\mathcal{L}_0$ the Fokker Planck operator given Eq.\ref{fokker_planck}.
The above equation obeyed by $\hat{G}_t$ is, under disguise, a wave propagation  equation, in the $x$-direction -although it lacks the diffusion term in $x$, such as in \cite{DerridaSpohn1988}-. This partial differential equation seems rather difficult to solve, but our analysis only requires the asymptotic speed of the wave, obtained from the exponential decay of the front $\gamma$. Plugging the Ansatz Eq.\ref{ansatz_bouchaud} and identifying terms of order $1$ and terms of order $e^{-\gamma(x-ct)}$, Eq.\ref{hatG_eq} reduces to the following system:
\bea
\label{first_order_eq}
0 &= \mathcal{L}_0 Q(\eta) \\
\nonumber
R c \gamma &=  \mathcal{L}_0 R + \gamma \eta R + \lambda (1/2)^\gamma Q \int \rmd\eta R(\eta) \\
\label{second_order_eq}
&  + R(\eta) \lambda ((1/2)^\gamma-1)
\eea

Eq.\ref{first_order_eq} states that $Q(\eta)$ is the stationary distribution of the process $\eta(t)$, consistently with Eq.\ref{asymp_ansatz}. The scaling degrees of freedom of both equations are fixed by the normalization contraints Eq.\ref{norm_const}. Once $Q$ is known, Eq.\ref{second_order_eq} is simply an inhomogeneous Sturm Liouville problem, whose operator is very similar to the Fokker-Planck operator of $\eta$, aside from the bias $\gamma \, \eta$, intimately related to the decay of the front.

It is convenient to introduce $Q_H(\eta) = \sqrt{Q(\eta)}= \mathcal{N}^{-1/2} \exp(-\Phi(\eta)/2)$. Transforming further $\mathcal{L}_0$ into an hermitian operator is accomplished by the change $R(\eta) = \lambda  2^{-\gamma} Q_H (\eta) S(\eta)$, and multiplying the whole equation by $ \sqrt{\mathcal{N}} \exp \left(\Phi (\eta)/2 \right)$ leaves us with the new system:
\bea
&\mathcal{L}_H S + \gamma \eta S  - \hat{\lambda} S (\eta)= - Q_H 
\label{SLversion}\\
 &\hat{\lambda} =  \lambda (1-(1/2)^\gamma) + c \gamma \nonumber \\
&\mathcal{L}_H = e^{\Phi /2} \mathcal{L}_0 e^{-\Phi /2} \nonumber
\eea
Solving this system is standard, and we adopt the Green Function (or resolvent) formalism \cite{economou1984green}. We first consider the homogeneous version of Eq.\ref{SLversion}:
\bea
\mathcal{L}_H S + \gamma \eta S + \alpha S  = 0 
\label{homo_SL}
\eea
for any real $\alpha$. As we assume a constant diffusion coefficient $D_2$, it is instructive to cast Eq.\ref{homo_SL} into a Schr\"odinger form:
\bea
&\frac{\partial ^2 S}{ \partial \eta ^2} = V(\eta)S - \frac{\alpha}{D_2} S   \\
&V(\eta)  = \frac{D_1(\eta)^2}{4 D_2^2} - \frac{D_1(\eta)' + 2 \gamma \eta}{2 D_2} \label{pot_schrodinger}
\eea
This analogy is particularly useful for exploring some exactly solvable models, as we can draw from the wisdom in Quantum Mechanics, and we will illustrate it through some examples below. The regular Sturm-Liouville theory asserts that solutions of Eq.\ref{homo_SL} can be decomposed over the eigenset $\lbrace \alpha_n \rbrace$ and $\lbrace \phi_n \rbrace$, $n \in \mathbb{N}^{+}$. We will assume from now on that Eq.\ref{homo_SL} has at least one bound state solution (in other words, at least $\alpha_0$ is isolated, at the bottom of the spectrum), the significance of such hypothesis will become clearer later on. We also use the common convention to write the decomposition as a discrete sum:
\bea
(\mathcal{L^H} + \gamma \eta) \phi_n = - \alpha_n \phi_n  \nonumber \\
S(\eta) = \sum _n s_n \phi_n (\eta)  \nonumber
\eea
Plugging it into Eq.\ref{SLversion}:
\bea
\sum_n (- \alpha_n - \hat{\lambda}) s_n \phi_n = -Q_H  \nonumber
\eea
and because $\lbrace \phi_n \rbrace$ is a complete orthonormal basis:
\bea
s_n (\alpha_n + \hat{\lambda}) = \langle Q_H | \phi_n \rangle  \nonumber
\eea
We can therefore write $S(\eta)$, and $R(\eta)$ decomposed as:
\bea
S(\eta) & = \sum_n \frac{\langle Q_H | \phi_n \rangle}{\alpha_n + \hat{\lambda}} \phi_n (\eta)  \nonumber \\
R(\eta)& = \lambda 2^{-\gamma} \sum_n \frac{\langle Q_H | \phi_n \rangle}{\alpha_n + \hat{\lambda}} Q_H(\eta) \phi_n (\eta)  \nonumber
\eea
Given proper boundary conditions, one can finally recover $c(\gamma)$ as an implicit equation by enforcing the self consistent condition $ \int d \eta\, R(\eta) =1$, leading to:
\bea
\frac{2^{\gamma}}{\lambda} & =  \sum_n \frac{\langle Q_H | \phi_n \rangle}{\alpha_n + \hat{\lambda}} \int_{\eta} d\eta Q_H(\eta) \phi_n (\eta) \nonumber \\
 & =  \sum_n \frac{\langle Q_H | \phi_n \rangle^2}{\alpha_n + \hat{\lambda}} \nonumber
\eea
The quantity:
\bea
G_{\gamma}(z) = \sum_n \frac{\langle \phi_n | \phi_n \rangle}{\alpha_n - z} \nonumber
\eea
is known as the resolvent operator. $G_0(z)$ corresponds to the Green function of the system with no bias $\gamma$, and its lowest eigenvector is precisely $Q_H(\eta)$, of eigenvalue $\alpha_0 =0$. We can compactly rewrite the above system as:
\begin{align}
\langle Q_H | G_{\gamma} \left( \lambda(2^{-\gamma}-1) - c \, \gamma \right) | Q_H \rangle = \frac{2^{\gamma}}{\lambda}
\label{final_res}
\end{align}
The above formula allows to extract $c$ as a function of the front decay $\gamma$. But it requires a rather suble analysis of the behaviour of the frond decay in travelling wave equations \cite{DerridaSpohn1988}. We recall this analysis in our setup in the following. 

\begin{figure}[hbt!]
\includegraphics[width=.43\textwidth, trim= 1cm 0 0 0]{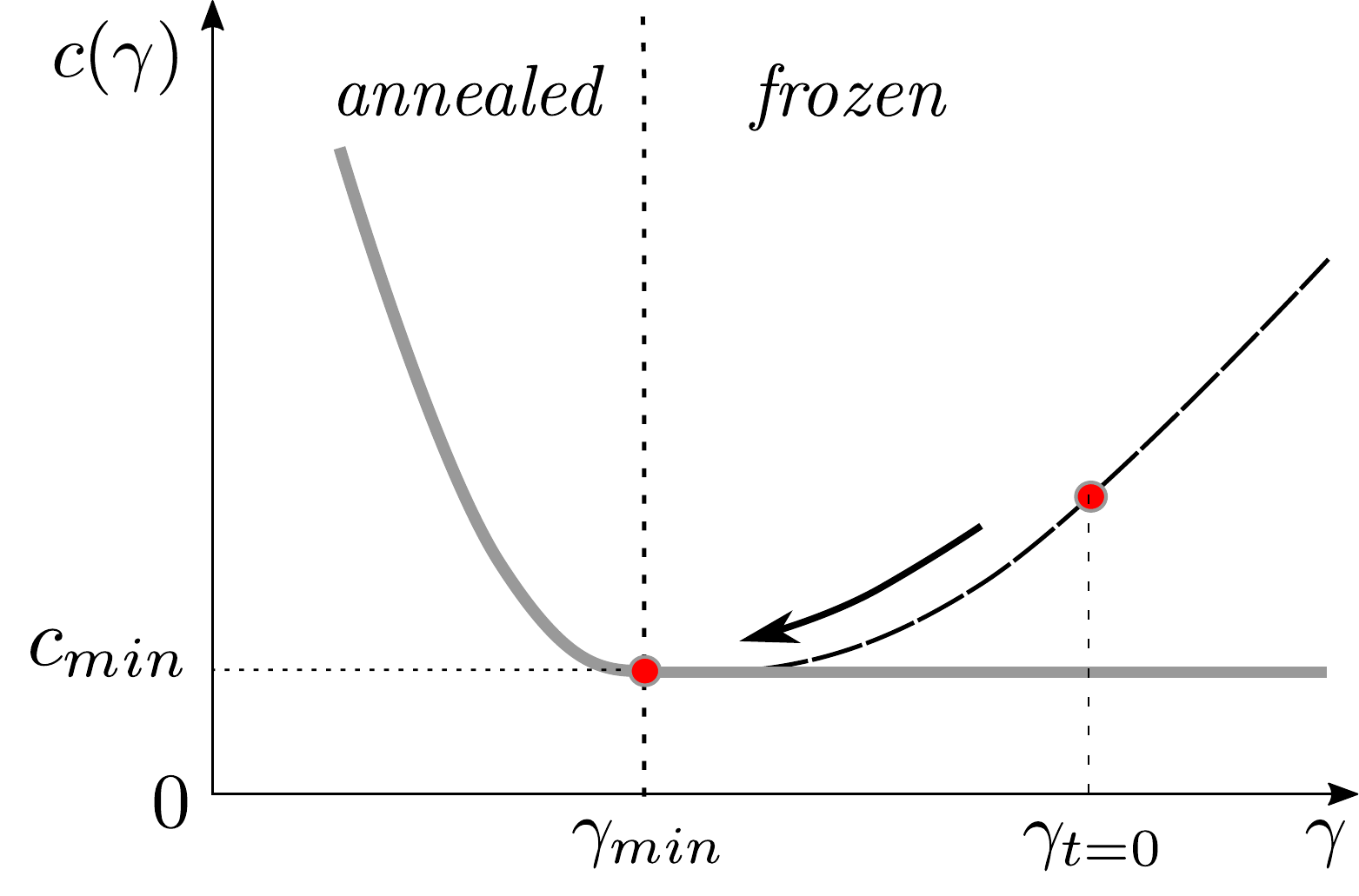} 
\caption{\textbf{(Sketch) The growth rate $c$ as a function of $\gamma$}. Typically, $c(\gamma)$ decreases at small $\gamma$ and increases at large $\gamma$. The increasing branch is not realized and any front initially prepared with a decay on this branch $\gamma_{t=0}> \gamma_{min} $ will relax towards the minimum, asymptotically propagating at a speed $c_{min}$ (the frozen regime).}\label{sketch_cgamma}
\end{figure}

\section{General relations between growth and diffusion \label{general_sect}}

\subsection{The front relaxation}

As written, Eq.\ref{final_res} is an implicit relation between $c$ and the decay of the front $\gamma$, with the parameters of the noise $D_1$, $D_2$ and the diffusion rate $\lambda$ as parameters. Generically, the curve $c(\gamma)$ exhibits a minimum at $\gamma_{min}$: for a range of $c$, the variable $\gamma$ is double valued (see Fig.\ref{sketch_cgamma}). In the wave propagation literature \cite{DerridaSpohn1988,bramson1983convergence}, it is known that, in fact, the increasing branch $\gamma > \gamma_{min}$ is never realized and the following mechanism takes place: when a front is prepared with a sharper decay than $\gamma_{min}$, this decay relaxes over time towards $\gamma_{min}$. Given Eq.\ref{init_cond}, the front is initially prepared with a decay $\gamma_0 = 1$. Two cases are possible:
\begin{itemize}
\item If $\gamma_{min} >1$, the propagation of the wave with $\gamma_{t=0}=1$ is possible: such situation corresponds to the \textit{annealed regime}, and occurs, for example, at large diffusion $\lambda$. Plugging $\gamma =1$ in Eq.\ref{final_res} leaves us with $c$ as a implicit function of the noise and $\lambda$. We refer to this portion of the curve as the \textit{annealed branch}.

\item Instead, if $\gamma_{min} <1$, the front broadens towards $\gamma_{min}$, the \textit{frozen regime} (see Fig.\ref{sketch_cgamma}). Then $\gamma$ is asymptotically fixed to $\gamma_{min}$ (itself a function of the parameters) and plugging its value in Eq.\ref{final_res} gives back $c$ as a function of $\lambda$. We refer to this portion of the curve as the \textit{quenched branch}.
\end{itemize}

\begin{figure}[hbt!]
\centering
\includegraphics[width=.40\textwidth, trim= 1cm 0 0 0]{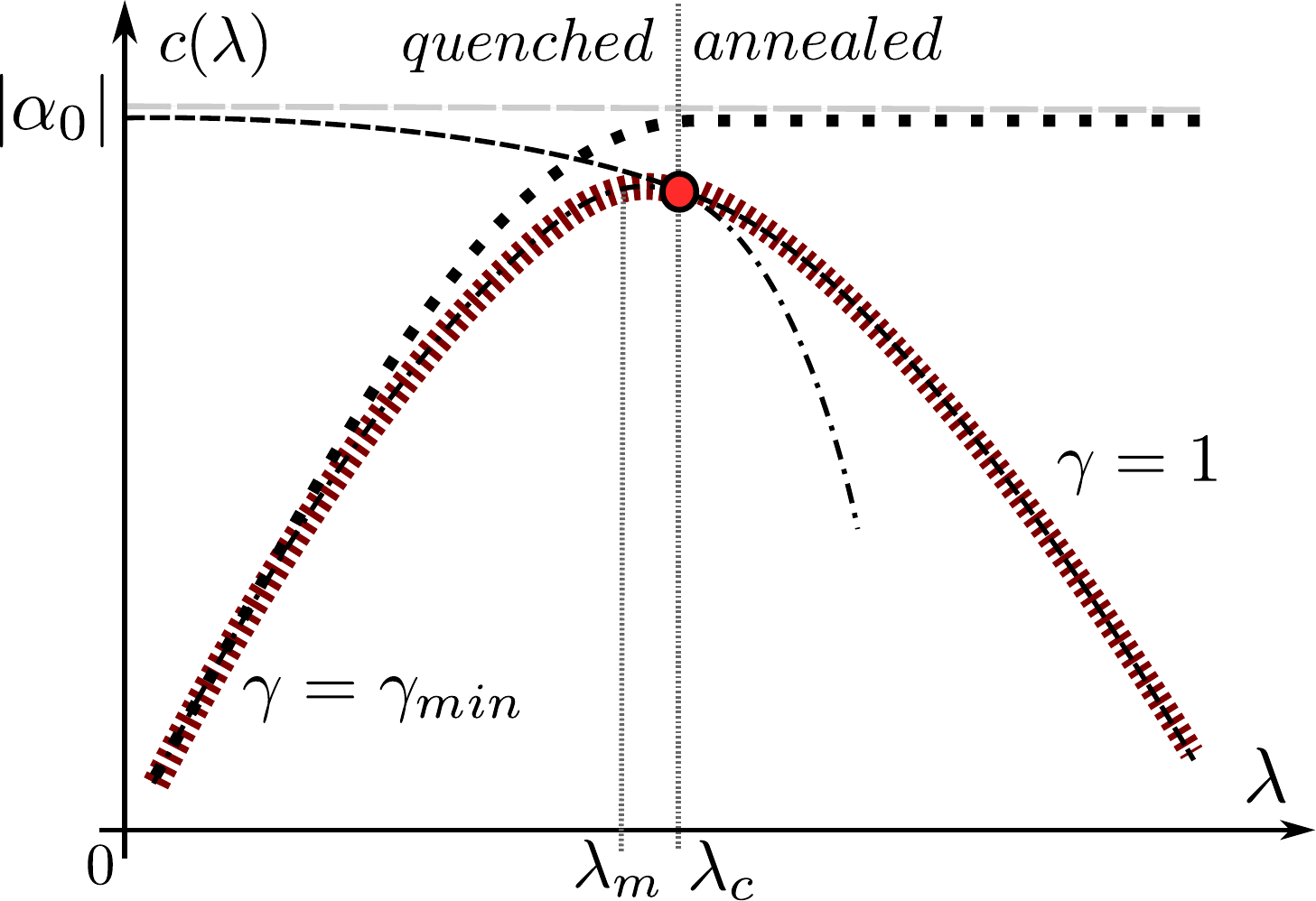} 
\caption{\textbf{(Sketch) The growth rate $c$ as a function of $\lambda$, in log-log plot}. Once the critical diffusion rate $\lambda_c$ is fixed by $\gamma_{min}=1$, it separates two regimes corresponding to quenched (dashed-dot) and annealed (dahsed) branch solutions of Eq.\ref{final_res}, depending on the value of $\gamma$. Both curves touch at $\lambda_c$. The correct curve is depicted in large, red, dashes for noises with a non-zero correlation time. The white-noise case is also depicted, with black squares, and reaches a plateau equal to $|\alpha_0 (\gamma=1)|$ at $\lambda_c$. Typical growth rates exhibit a maximum at a value $\lambda_m$, in the quenched phase $\lambda_m < \lambda_c$.}\label{sketch_clambda}
\end{figure}

When $\gamma_{min}=1$, the parameters are tuned right on the freezing transition, and we denote $\lambda_c$ the critical diffusion rate. The typical shape of the whole curve is shown on the sketch Fig.\ref{sketch_clambda}. The junction of both branches is therefore at $\lambda_c$ and they also have a global maximum at $\lambda_m$. The remaining part of the paper will be dedicated to the analysis of such curve for general processes, and illustrated on particular examples. 

As described by Eq.\ref{final_res}, to obtain $c(\lambda)$, one merely needs to obtain the resolvent of Eq.\ref{homo_SL}, the same as the resolvent of the operator $\mathcal{L_H}$ with an additional linear bias of amplitude $\gamma \eta$. Green functions are usually difficult to compute and such task is not easy. This very problem is nonetheless not new and has triggered a large activity in the somewhat unrelated field of Quantum Mechanics (QM), under the name of \textit{Stark effect} \cite{schiff1955quantum}: how is a bounded electron perturbed when an electric field is switched on? Of course, the mapping from It\=o process to Schr\"odinger potential may sometimes lead to complicate expressions of $f(\eta)$, but it also provides a way to leverage the computational means developed to tackle the Stark effect. Let us illustrate the similarity of both problems.

Note that $\hat{\lambda}>0$ for any $\gamma$, and consider first the case $\gamma$ very small. In the right side of Eq.\ref{final_res}, all the terms in the sum, except for $n=0$, are close to $0$, due to the vanishing overlaps. Hence, Eq.\ref{final_res} reduces to good approximation to:
\begin{align}
\frac{2^{\gamma}}{\lambda} \simeq \frac{ \langle Q_H |\phi_0 \rangle^2}{\alpha_0(\gamma) + \hat{\lambda}} \label{approx_green}
\end{align}
from which we will extract the asymptotics for $\gamma \rightarrow 0$. At $\gamma=0$, excited states all have a higher positive energy, and $\hat{\lambda}>0$. Once $\gamma$ differs from $0$, $\alpha_0(\gamma)$ necessarily becomes negative, a well-known result in QM \cite{schiff1955quantum}. As $\gamma$ goes to $1$, the behaviour of the series $\lbrace \alpha_n \rbrace_n$ strongly depends on the details of the disorder, and some eigenvalues may cross the $y$-axis, also becoming negative. Therefore, many branches of solutions of Eq.\ref{final_res} appear, but because we expect $c(\gamma)$ continuous, the physical solution remains close to the pole at $\alpha_0$, and so $c(\gamma) < |\alpha_0(\gamma)|$ for any $\gamma$.

An important quantity, usually coined the \textit{polarisability} $\epsilon$, is defined as:
\begin{align}
\alpha_0 (\gamma) = - \epsilon \gamma ^2 + O(\gamma ^3)
\label{polarisation_eq}
\end{align} 
The vanishing of the first order term stems from the fact that the mean of $\eta$ is set to $0$. The value of $\epsilon$ is obtained either using the Rayleigh-Schrodinger theory, or simply expanding the stationary probability distribution $Q(\eta)=|Q_H(\eta)|^2$ in small $\gamma$, obtaining for the energy at second order, after some manipulations:
\begin{align}
\epsilon &= \frac{1}{D_2} \int_{-\infty} ^{\infty} \frac{dx}{Q(x)}\left(\int_{-\infty}^{x} s Q(s) ds \right)^2 \nonumber\\
& = \langle \eta^2 \rangle_Q \, T
\label{jung_risken_rel}
\end{align}
We have also used the general expression for the correlation time defined in Eq.\ref{risken-jung}. This is an example of Green-Kubo identity: $T$ is obtained by integrating the temporal two point function $\langle \eta(0) \eta(t) \rangle$, whereas $\epsilon$ describes the response to a linear forcing proportional to $\gamma$. In our context, $\epsilon$ quantifies the propensity of $\eta$ to ``yield'' under the effect of the bias $\gamma$. The characteristics of a soft -or very sensitive to the biais- process become rather clear from Eq.\ref{jung_risken_rel}: it should widely fluctuate or be long time-correlated. 

Note that Eq.\ref{approx_green} is, in many cases, a suitable approximation also for large $\gamma$. Indeed, as $\gamma$ grows, the left side of Eq.\ref{final_res} blows up exponentially, forcing $\hat{\lambda}$ to concentrate around $\alpha_0$. This ultimately depends on the asymptotic behaviour of $\lbrace \alpha_n \rbrace$. We therefore turn onto a more detailed study of the asymptotics of both annealed and quenched branches.

%

\subsection{The annealed branch}

We first consider the behaviour of $c_a(\lambda)$ with $\gamma=1$ in the limit of weak diffusion $\lambda \rightarrow 0$. It provides a useful upper bound for $c_q(\lambda)$:
\begin{align}
c_{a}(\lambda) &=|\alpha_0(\gamma=1)| -  \nonumber \\
&\left(1- \langle Q_H | \phi_0 \rangle^2_{\gamma=1}\right)\frac{\lambda}{2} + O(\lambda ^2)
\label{zerodiff_exp}
\end{align}

Knowing that $\alpha_0(\gamma = 0) = 0$, $|\alpha_0(\gamma=1)|$ again measures the ability of the noise to "polarize" under the field $\gamma = 1$. The extreme case of a "stiff" process is the constant Langevin noise, for which a variation of $\lambda$ has no effect at all. Interestingly, Eq.\ref{zerodiff_exp} also provides a compact way to compute the Laplace transform of integrated Markov processes $\langle \exp(\int_0 ^t \eta(t) dt) \rangle$, an important endeavour in finance \cite{yor2012exponential}.

The large $\lambda$ limit requires to expand the right hand side of Eq.\ref{final_res} in inverse powers of $\lambda$ (to lighten the notations, all the overlaps and eigenvalues in the remaining of this subsection are evaluated at $\gamma = 1$), assuming the overlaps decay exponentially fast at large $i$:
\begin{align}
1 &=\sum_i \frac{\langle Q_H | \phi_i \rangle^2}{2 \alpha_i/\lambda + 1 + 2 c(\lambda)/\lambda}  \nonumber \\
1 &= \sum_i \langle Q_H | \phi_i \rangle^2 \left(1 -2 \frac{c(\lambda) + \alpha_i}{\lambda} +\cdots \right)  \nonumber
\end{align}
Using together the normalisation of $Q_H$, and the fact that $Q(\eta)$ has zero mean, hence $\langle E \rangle =
\sum_i \alpha_i \langle Q_H | \phi_i \rangle^2=  \langle Q_H |\hat{x} | Q_H \rangle=0$, we obtain:
\begin{align}
c(\lambda) = 2 &\frac{\sum_i \alpha_i ^2 \langle Q_H | \phi_i \rangle^2}{\lambda}  \nonumber\\
 &- 4 \frac{\sum_i \alpha_i ^3 \langle Q_H | \phi_i \rangle^2}{\lambda^2}+ O(\lambda^{-3})
\end{align}

The coefficient of the dominant decay can be rewritten using:
\begin{align}
\sum_i \alpha_i ^2 \langle Q_H | \phi_i \rangle^2 &= \int_{\eta} Q_H(\eta) (\mathcal{L}_H +\eta)^2 Q_H(\eta)  \nonumber\\
&= \int_{\eta} \eta^2 Q_H(\eta)^2 = \langle \eta^2 \rangle_{Q}
\label{scaling_largegamma}
\end{align}
Hence the dominant decay is given by the variance of $Q$. Higher order terms include higher moments of $Q$ and can be systematically computed.

\subsection{The quenched branch}

The quenched branch is more difficult to investigate, as one has also to obtain the location of the minimum $\gamma_{min}$ of $c(\gamma)$. We extract the small $\lambda$ expansion, obtained by considering Eq.\ref{approx_green}, under the assumption that both $\gamma_{min}$ and $c(\gamma_{min})$ go to $0$ as $\lambda \rightarrow 0$.
Plugging Eq.\ref{polarisation_eq} into Eq.\ref{approx_green} and balancing all the terms, we obtain:
\begin{align}
\gamma_{min} &= \sqrt{\frac{\lambda}{\epsilon}} + O(\lambda) \\
c(\gamma_{min}) &= 2 \sqrt{\epsilon \lambda} + O(\lambda) 
\label{scaling_smallgamma}
\end{align}
 
By expanding to higher order the lowest eigenvalues and overlaps of the resolvent, the approximation can be systematically improved, but the computation quickly becomes tedious. 

Note that both asymptotics, large and small $\lambda$, depend on the variance and the time correlation of the noise only, a manifestation of universality. One can shed light on those scalings \ref{scaling_largegamma} and Eq.\ref{scaling_smallgamma} using more hand-waving arguments and the Feynman-Kac representation \cite{rogers2000diffusions}:
\begin{align}
Z(x,t) = \Bigg \langle \exp \left(\int_0 \eta_X(s)(t-s) ds \right) \Bigg \rangle_{\pi_X}  \nonumber
\end{align}
where $X$ is a Poisson process over the space of sites, of rate $\lambda$ and distribution $\pi_X$. 

We consider first the small diffusion $\lambda$ case. Over a total time $t$, $\lambda t$ jumps occur, breaking $\int_0 \eta_X(s)(t-s) ds$ into $\lambda t$ pieces. Each of those pieces is the integral, over a time $1/\lambda$, of a time $T$-correlated noise, and so has a typical amplitude of $\sqrt{\langle \eta^2 \rangle_Q T t/\lambda}$. Deep in the quenched phase, the measure is dominated by the maximum over $X$, being roughly estimated by:
\begin{align}
\log(Z) \sim \lambda t \times \sqrt{\langle \eta^2 \rangle_Q T/\lambda}  \nonumber \\
c(\lambda) \simeq \log(Z)/t \sim \sqrt{\langle \eta^2 \rangle_Q T \lambda}  \nonumber
\end{align}

The high-$\lambda$ limit goes along similar same lines and has been presented in \cite{gueudre2014explore} in a different form: first recall that, for the white noise model, the free energy in the annealed phase is fixed to $\langle \eta^2 \rangle_Q$. At finite $T$ and large $\lambda$, the random walk is so fast, compared with $T$, that it only sees a frozen disorder on each site, before jumping onto another. Again $\int_0 \eta_X(s)(t-s) ds$ breaks into $\lambda t$ pieces, but each is now simply the integration, over a time $1/\lambda$, of a frozen random variable $\eta$, independently drawn from $Q(\eta)$. Therefore in this case:
\begin{align}
\log(Z) & \sim \lambda t \times \langle \eta^2 \rangle_Q/\lambda^2 \nonumber \\
c(\lambda) & \sim \langle \eta^2 \rangle_Q/\lambda \nonumber
\end{align}

\section{Particular processes \label{exactly_solvable}}

In this section, we illustrate the computational aspect of the approach, first solving the case of the Ornstein Ulhenbeck by an alternative, but equivalent, route to the one presented in \cite{gueudre2014explore}. We then go onto processes of bounded support, or with varying tails in their stationary distributions. Other solvable examples could be inspired by the literature on Stark effect \cite{robinett2009stark,angel1968hyperfine,bastard1983variational}.

\begin{figure*}
\begin{minipage}[b]{0.49\textwidth}
\includegraphics[width=\textwidth]{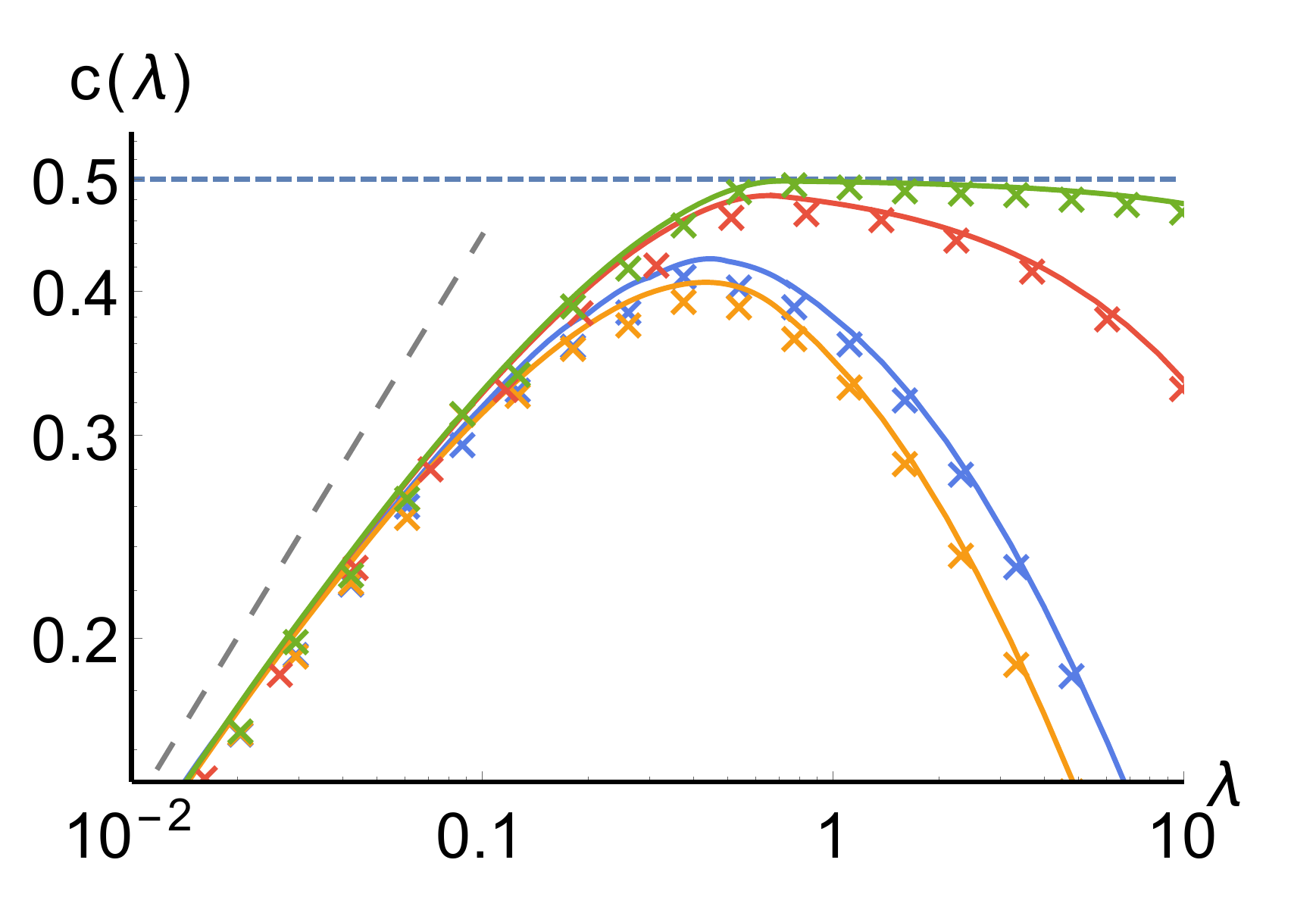} 
\caption{\textbf{$c(\lambda)$ as a function of $\lambda$ for an Ornstein-Ulhenbeck process}, with the set of parameters (from top to bottom): $D_2=5000$, $k=100$;  $D_2=50$, $k=10$; $D_2=1$, $k=\sqrt{2}$; $D_2=0.5$, $k=1$. The dashed line is the expansion Eq.\ref{scaling_smallgamma}. The scaling have been chosen so that the upper bound $|\alpha_0(\gamma=0)|$ is fixed to the value $0.5$. The numerics are performed on a system of size $N=10^6$ sites, up to a time $t_{tot}=500$, with the discretization time step $dt=0.001$.}
\label{ornstein_data}
\end{minipage}\hfill
\begin{minipage}[b]{0.49\textwidth}
\includegraphics[width=\textwidth]{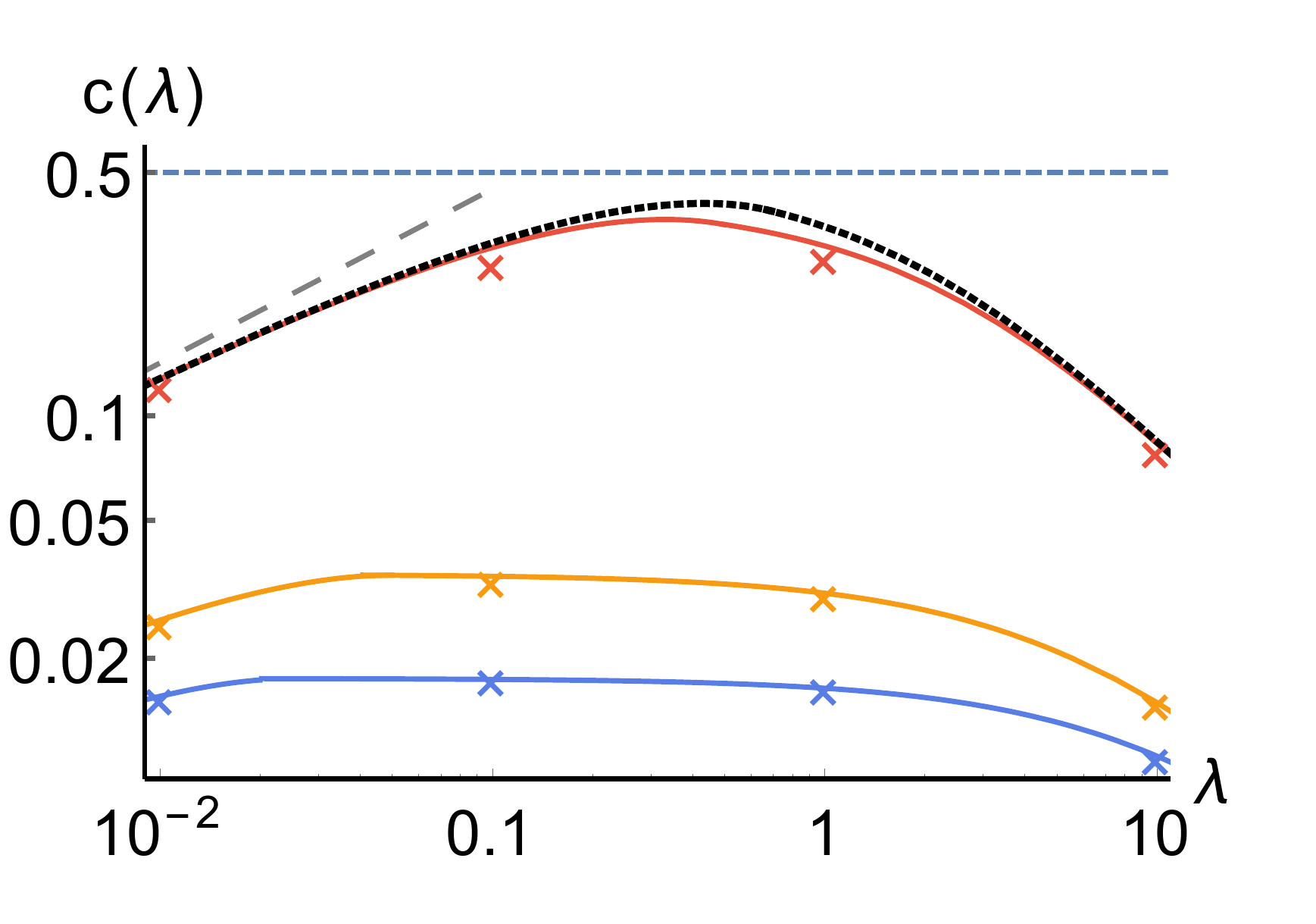} 
\caption{\textbf{$c(\lambda)$ as a function of $\lambda$ for the bounded process}, with the set of parameters (from top to bottom): $D_2$ and $a$ fitted from Eq.\ref{fitted};  $D_2=0.5$, $a=1$; $D_2=1$, $a=1$. The dotted line is a fit obtained from the OU process, matching the asymptotic behaviour. The dashed line is the expansion Eq.\ref{scaling_smallgamma}. The numerics are performed on a system of size $N=10^6$ sites, up to a time $t_{tot}=500$, with the discretization time step $dt=0.001$.}
\label{bounded_data}
\end{minipage}
\end{figure*}

\subsection{The Ornstein-Ulhenbeck process}

The Ornstein-Ulhenbeck (OU) process was the first colored generalization made \cite{gueudre2014explore}, to our knowledge. This corresponds in QM, to the harmonic oscillator, whose solution is completely known. The Fokker-Planck operator and stationary solution are:
\begin{align}
\mathcal{L}_0 &= D_2 \frac{\partial ^2}{\partial \eta ^2} + k \frac{\partial}{\partial \eta} \eta  \nonumber \\
f(\eta) &= \frac{k \eta^2}{2}  \nonumber \\
Q(\eta)&= \sqrt{\frac{k}{2 \pi D_2}} e^{-\frac{k \eta ^2}{2 D_2}}  \nonumber \\
Q_H (\eta) &= \left(\frac{k}{2 \pi D_2} \right)^{1/4} e^{-\frac{k \eta ^2}{4 D_2}}  \nonumber
\end{align}

The problem is equivalent to solving the Schrodinger equation in a potential given by Eq.\ref{pot_schrodinger}:
\bea
V(\eta) &= \frac{k^2 \eta^2}{4 D_2^2}-\frac{k + 2 \gamma \eta}{ 2 D_2} \nonumber
\eea
a tilted harmonic potential. Using $\tilde{\eta}=\sqrt{\frac{k}{2 D_2}} \eta - \frac{\sqrt{2 D_2}}{k^{3/2}} \gamma$, we reduce it to:
\bea
\frac{\partial ^2 S}{\partial \tilde{\eta} ^2} = (\tilde{\eta}^2 - \epsilon) S \nonumber \\
\epsilon = 1+ \frac{2 \alpha}{k} + \frac{2 D_2 \gamma ^2}{k^3} \nonumber
\eea
The propagator of the Harmonic oscillator goes by the name of the $\textit{Mehler formula}$. In the $(\tilde{\eta},\epsilon)$ set of variables:
\bea
&K(\tilde{\eta}_1,\tilde{\eta}_2,t) = \frac{1}{\sqrt{2 \pi \sinh (2t)}} \times \nonumber \\
&\exp (  \coth(2 t) (\tilde{\eta}_1^2 +\tilde{\eta}_2^2)/2 + \text{cosech}(2 t) \tilde{\eta}_1 \tilde{\eta}_2 ) \nonumber
\eea
The resolvent is simply the Laplace transform of the propagator $K(\tilde{\eta}_1,\tilde{\eta}_2,t)$ with respect to $t$:
\bea
\frac{2^{\gamma}}{\lambda} =\int_{\tilde{\eta}_1,\tilde{\eta}_2} d\tilde{\eta}_1 d\tilde{\eta}_2 \int_{t=0} ^{\infty} dt e^{-\hat{\lambda} t} K(\tilde{\eta}_1, \tilde{\eta}_2,t) \nonumber
\eea
Performing both gaussian integrals in $\tilde{\eta}_1$ and $\tilde{\eta}_2$, we are left with:
\bea
\frac{2^{\gamma}}{\lambda} =  \int_0 ^{\infty} dt \exp \left(\frac{D_2 \gamma^2}{k^3} (e^{-t k}-1) + t (\frac{\gamma^2 D_2}{k^2} - \hat{\lambda}) \right) 
\label{final_res_OU}
\eea
Another route (detailed in Appendix \ref{prl_OU}) is to fully diagonalize $\mathcal{L_H}$ and write down the resolvent as an infinite sum. A numerical confirmation of the above result is plotted Fig.\ref{ornstein_data}. The upper bound is $|\alpha_0(\gamma=1)| = D_2/k^2$, and the set of parameters in Fig.\ref{ornstein_data} has been chosen so that this upper bound is fixed to $1/2$. As $T = 1/k$ tends to $0$ (the white noise limit), $c(\lambda)$ saturates at the plateau $c(\lambda) = |\alpha_0(\gamma=1)| $ in the annealed phase. This limit is singular however, as for any small $T>0$, $c(\lambda)$ decays as $\lambda ^{-1}$. 

\subsection{The bounded noise} 

Another case of common interest, especially in condensed matter, is the noise of bounded support. It corresponds to a particle in an infinite well, submitted to a uniform electric field, and is again solvable \cite{merzbacher1970quantum}, although we end up with a set of transcendental equations.

To simplify slightly the analysis, we set $V(\eta)$ to be a square infinite well, which translates into a bounded but rather contrived It\={o} process. At $\gamma = 0$, we have:
\bea
V_0(\eta) &= - \frac{\pi^2}{4 a^2} \text{	for 	} |\eta | < a \nonumber\\
f(\eta)&= - 2 D_2 \ln \left(\cos(\frac{\pi \eta}{2 a}) \right) \nonumber\\
Q(\eta) &= a^{-1} \cos^2 \left( \frac{\pi \eta}{2 a} \right)\nonumber\\
Q_H(\eta) &= a^{-1/2} \cos \left( \frac{\pi \eta}{2 a} \right)\nonumber
\eea
The eigenset is simply made of Airy functions. The potential with bias is $V(\eta) = -\pi^2/(4 a^2) - \gamma \eta/D_2$ and after the change of variables:
\bea
\tilde{\eta} = -\left( \frac{\gamma}{D_2} \right) ^{1/3} \left(\eta + \frac{\alpha}{\gamma} +\frac{\pi ^2 D_2}{4 a^2 \gamma} \right) \nonumber
\eea
we obtain the following eigenbasis, with their according boundary conditions:
\bea
&\phi_n(\eta) = a_n Ai (\tilde{\eta}) + b_n Bi (\tilde{\eta}) \nonumber \\
&\tilde{\eta}^b_{\pm}= -\left( \frac{\gamma}{D_2} \right) ^{1/3} \left(\pm a + \frac{\alpha}{\gamma} +\frac{\pi ^2 D_2}{4 a^2 \gamma} \right) \nonumber \\
&Ai(\tilde{\eta}^b_{+})Bi(\tilde{\eta}^b_{-})=Ai(\tilde{\eta}^b_{-})Bi(\tilde{\eta}^b_{+}) \nonumber
\eea
The discrete eigenvalues $\lbrace \alpha_n \rbrace_n$ are solutions of the above transcendental equation. Once this discrete set of eigenvalues is determined, $a_n$ and $b_n$ can be fixed so that the set $\phi_n$ is normalized and obeys the boundary conditions. We compute the first $N=10$ terms of the resolvent as an estimate. $c$ as a function of $\gamma$ is plotted Fig.\ref{bounded_data} and compared with numerical simulations. Once again, the agreement is excellent. On Fig.\ref{bounded_data}, we also have compared this bounded process with an Ornstein-Ulhenbeck one, matching both $T$ and $\langle \eta^2 \rangle$, which read:
\begin{align}
\langle \eta^2 \rangle_Q = \frac{a^2(1 -6/\pi^2)}{3} = \frac{1}{2} \nonumber\\
T = \frac{a^2(15/\pi^2-1)}{D_2(\pi^2-6)} = 1 \label{fitted}
\end{align}

Both curves are quite similar, the largest deviation occurs around the freezing transition. It emphasizes the difficulty of choosing a faithful modelling of systems that sit around $\lambda_c$.

\subsection{The role of the tails}

Growth processes can be seen as extremal in some sense: their statistics are dominated by those space-time paths that manage to collect the largest amount of resources. Inspired by the theory of extreme statistics, one would expect the tails of $Q(\eta)$ to play a prevalent role. The asymptotics mentioned in Section \ref{general_sect} only depend on $\langle \eta^2 \rangle_Q$ and $T$. To analyse the effects of the tail of $D_1(\eta)$ on $T$, for example, we define the famility distribution $Q_{\mu}$ obtained from $D_1(\eta) \sim sign(\eta) \, |\eta|^{\mu}$, such that $D_2$ and $\langle \eta^2 \rangle_{Q_{\mu}}$ are normalized to $1$. This family smoothly interpolates from the harmonic potential $\mu=1$ to the infinite well $\mu=\infty$. We then compute $\epsilon(\mu)$ from Eq.\ref{jung_risken_rel} using the expression of $Q_{\mu}$. It turns out that $\epsilon(\mu)$ has a minimum at $\mu=1$ (the case of the OU process $\epsilon(1) = 1$) and tends to $6/5$ at infinity (the process with a uniform stationary distribution and unit variance). The conclusion is that, \textit{at fixed variance}, thinner tails yield an enhanced growth. Although somewhat conterintuitive, it can be traced to the flatter nature of the potential $\Phi(\eta)$ at large $\mu$, when $\eta$ is close to $0$, increasing the polarizability of $\eta$. 

The perturbative results in the range $\mu \in (0,1)$ have to be taken with a grain of salt, as the perturbation becomes singular and requires a more elaborate treatment \cite{bender1999advanced}. For example, at $\mu=0$, the process has a Laplace stationary distribution, for which an exact solution exists and one can show that the energy gap between $\alpha_0$ and the rest of the spectrum vanishes even at small $\gamma$. We return to it later, when examining the limitations of this spectral approach.

\section{Discussion \label{discussion}}

The previous exactly solvable cases and the expansions make all the more obvious the existence -and robustness- of both a freezing transition point $\lambda_c$, and a maximum at $\lambda_m$ in the growth rate. 
At diffusion low enough, the total population is not a self-averaging quantity, and so $c_q < c_{a} $. The gap between $c_q$ and $c_{a}$ is due to heavy tails and strong correlations between the $Z_i(t)$. Those factors grow as diffusion decreases, favorizing condensation onto few sites. At $\lambda=0$, $Z_i$ merely reduces to the exponential of $\int_t dt \eta(t)$, the integrated It\={o} process: $\log Z$ is essentially a Gaussian of zero average and growing variance $\langle \eta^2 \rangle T t$, and $Z$, a log-normal, heavy-tailed distribution.  

There seems to be no close formula neither for the value $\lambda_m$ at which the freezing transition occurs, nor for the point of optimal growth $\lambda_c$. Nonetheless, for any process $\eta$, the annealed branch $c_a(\lambda)$ is monotonically decreasing with $\lambda$ (see Appendix \ref{monotonous} for a proof), and $c_a(0)=|\alpha_0| \simeq \epsilon$. Assuming the quenched branch is differentiable, we deduce that necessarily $\lambda_m \leq \lambda_c$, with equality in the limiting case of white noise. The fact that the optimum always lays in the quenched phase is intriguing, and reminiscent of the Zipf law, a very general attempt to explain the predominance of power-laws in natural systems. The present case falls in the category of \textit{highly optimized tolerance} \cite{sornette2006critical}: when optimized, complex systems have a tendency to develop algebraic tails and experimental studies have shown that they are found close from the optimal point \cite{caraglio2016export}. Given that $\lambda_m$ and $\lambda_c$ are not far, it also shows how systems poised at optimality could deceiptively look critical \cite{newman1996self,mora2011biological}.

\begin{figure}[hbt!]
\centering
\includegraphics[width=.50\textwidth, trim= 0 0 0 0]{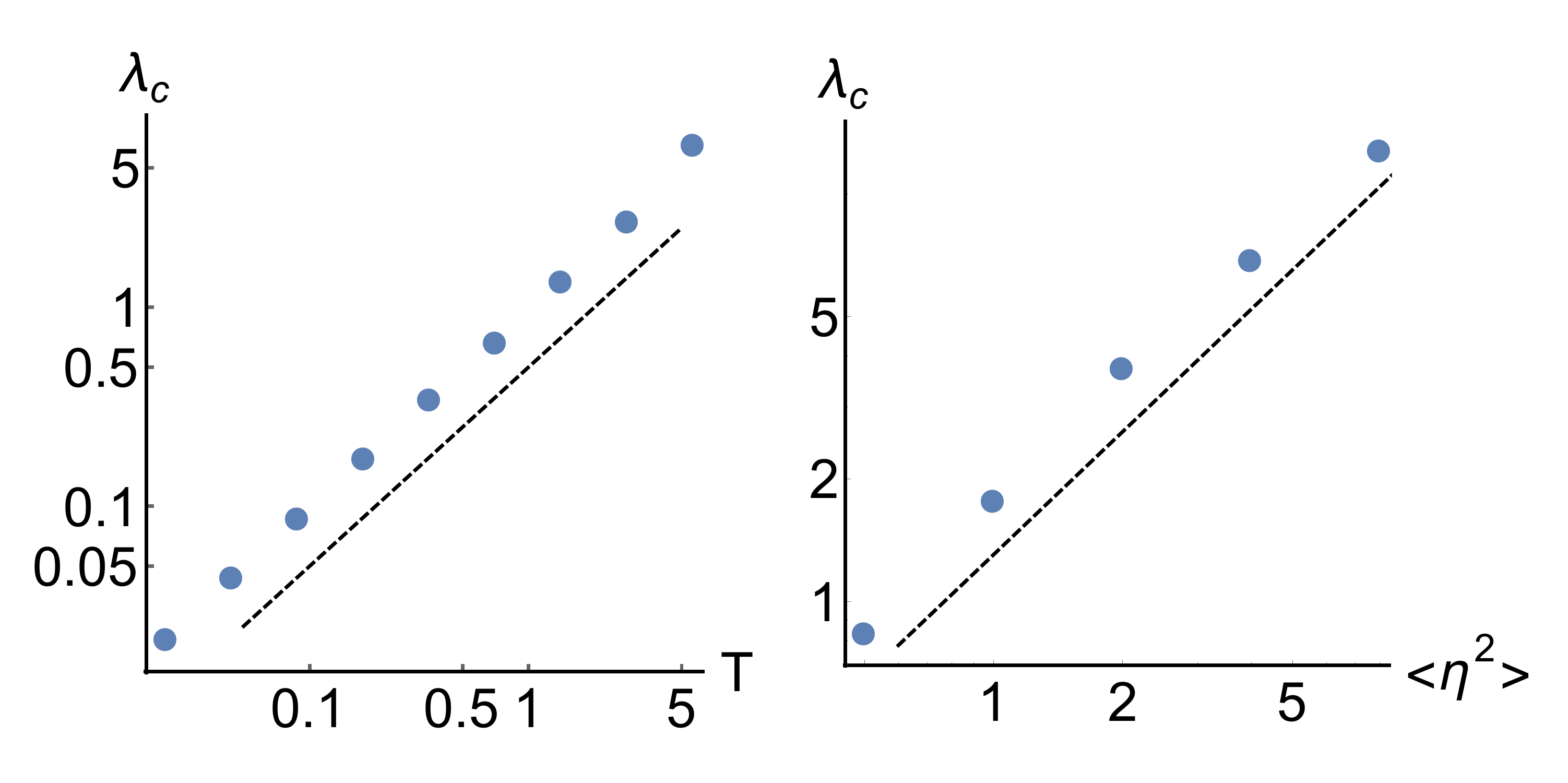} 
\caption{\textbf{Scaling of $\lambda_c$ for the OU process} (Left) Scaling of $\lambda_c$ with $T$, and $\langle \eta^2 \rangle = 1/\sqrt{2}$. (Right)  Scaling of $\lambda_c$ with $\langle \eta^2 \rangle$, and $T = 1.0$. Both dashed lines are guidelines of unit slopes.}\label{scaling_lambdam}
\end{figure}

\begin{figure}[hbt!]
\centering
\includegraphics[width=.45\textwidth, trim= 0 0 0 0]{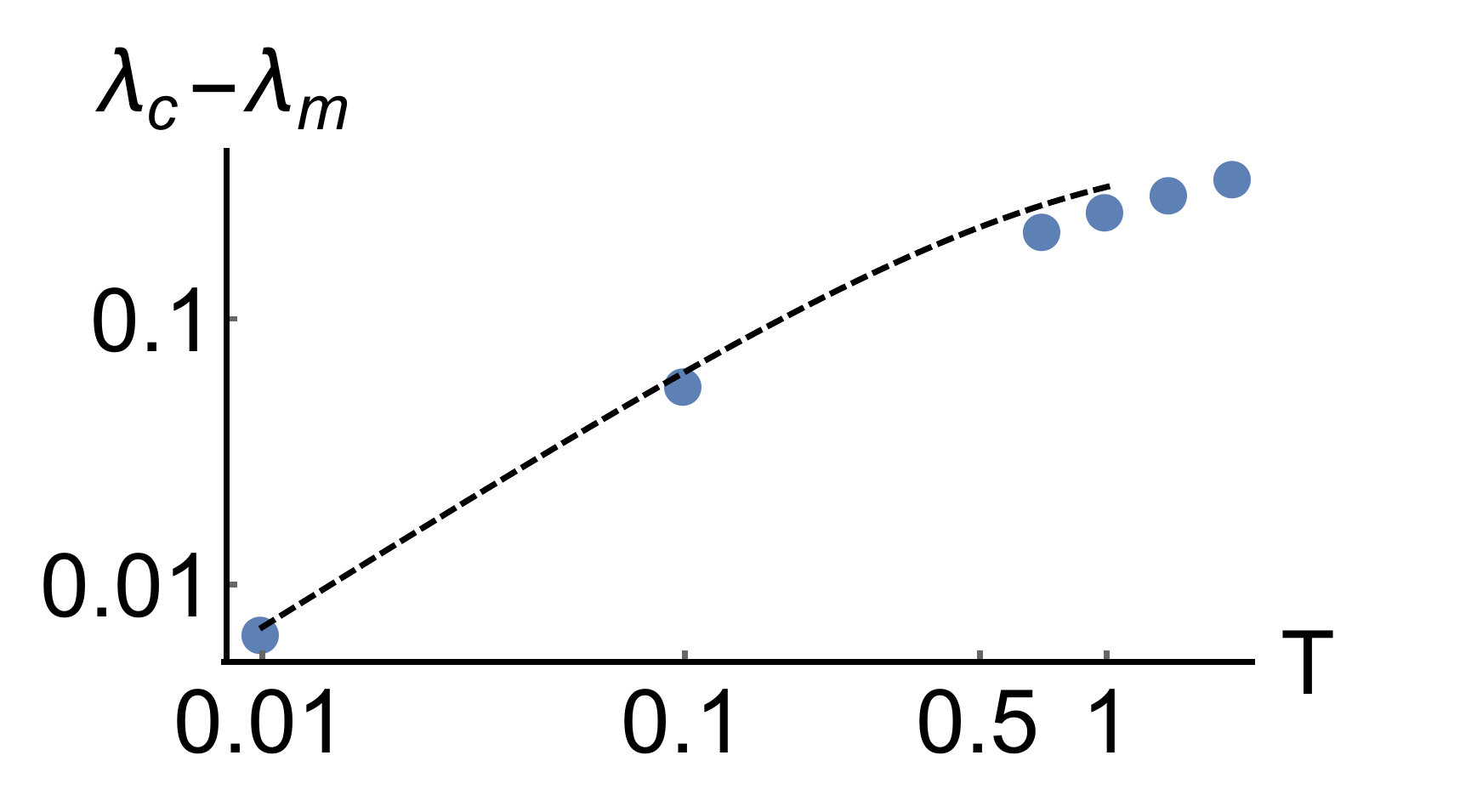} 
\caption{\textbf{Scaling of the difference $\lambda_c - \lambda_m$ for the OU process}. We have fixed $\epsilon = \frac{D_2}{k^2} = \frac{1}{2}$. The blue dots are the result of the numerical solution of Eq.\ref{final_res_OU} for specific values of $D_2$ and $k=1/T$, while the black dashed line is a small $T$ expansion: $\lambda_c - \lambda_m$ widens linearly with $T$, close to $T=0$.}\label{scaling_gap}
\end{figure}

A rough estimation of the position of $\lambda_c$ (or $\lambda_m$) is obtained by balancing the asymptotics with the upper bound $c_a(\lambda=0)$, leading to:
\begin{align}
\lambda_c &\sim \lambda_m \sim \langle \eta^2 \rangle \times T
\label{scaling_lambda}
\end{align}
In Fig.\ref{scaling_lambdam}, we tested its validity by numerically solving Eq.\ref{final_res_OU} for the specific OU process, fixing either $\langle \eta^2 \rangle$ or $T$. It is also possible to investigate the behaviour of $\lambda_m$ and $\lambda_c$ close to the white noise limit $T \rightarrow 0$, at fixed $\epsilon = 1/2$ (the case presented in Fig.\ref{ornstein_data}). A tedious expansion at small $T$ from Eq.\ref{final_res_OU} yields both $\lambda_c(T) \simeq \frac{2 \epsilon}{\log 4 + T \epsilon(1-\log 2)}$ ($\lambda_c(0)=0.7213...$) and the fact that the gap $\lambda_c - \lambda_m$ grows linearly with $T$, with a complicated prefactor that we do not report. This confirms the scaling presented in Eq.\ref{scaling_gap}. In the regime of widely fluctuating noises $T \ll \langle \eta^2 \rangle$, $\lambda_m \simeq \lambda_c$ and a large plateau in $c(\lambda)$ develops around those transition points: the diffusion is still small enough for the noise to be seen as a quasi white noise. This suggests that optimal growth is more robust in a widly varying environment, a rather surprising finding.

On a more practical side, it is often difficult to characterize the properties of the miscrocopic noise $\eta(t)$, and only macroscopic observables are measured. Such situations are common occurences in biology for example, where concentrations of proteins or bacteries are much easier to obtain than levels of mRNA or nutrients they harvest. Within the present class of growth models, $\langle \eta^2 \rangle_Q$ and $T$ can be extracted from both small and large $\lambda$ (assuming $\lambda$ is a control parameter of the experiment). Those two values, the most salient features of $\eta$, are enough to fit one of the solvable models onto the experience at hand and estimate $\lambda_m$ and $\lambda_c$. We believe the mechanims presented above to be of more general scope and investigating both the interplay between $\lambda_m$ and $\lambda_c$, as well as their presence in other, non mean-field growth models, would be a worthy subject of investigation.

To conclude, we would like to comment on some limitations. The original case made in \cite{DerridaSpohn1988}, and most of the subsequent literature, concerns the pure white noise (also called \textit{branching Brownian Motion}). Its evolution cannot be cast into a well-defined It\={o} equation, but may be obtained as a rather singular limit with $T \rightarrow 0$. On the other hand, It\={o} processes with no stationary distribution $Q(\eta)$ -such as the Brownian Motion-, fall out of the present analysis. Yet we expect them to have no freezing transition: the wandering of those processes is so important that few branches of the tree, if not a single one, should always dominate the statistics, but a more precise study would also be welcomed.

The requirement of at least one isolated state at the bottom of the spectrum is a more subtle issue. In principle, such restriction is not necessary, although one would have to tackle the continuous part of the spectrum describing the extended states. The process with the stationary Laplace distribution $Q(\eta) \sim \exp(-|x|)$ is an enlightening example. It translates as a Dirac potential $V(\eta) \sim \delta(\eta)$ in Eq.\ref{pot_schrodinger}. It is known that a particle in such a narrow potential, and also submitted to an electric field, has no bound state, even for $\gamma$ infinitesimally small (see \cite{lukes1969exact} and Appendix \ref{app_exp_model} for more details). Therefore the resolvent has no simple pole, and the expansions presented in Section \ref{general_sect} are not valid anymore. One has to integrate over the branch cut of $G_{\gamma}$, which extends over the whole real axis, and regularize it with an $\epsilon$-prescription. While one can write down such equation (see Appendix \ref{app_exp_model}), the resulting integrand involves complex, oscillating, terms, that are very difficult to tackle numerically. In models of the same flavour (such as the Random Energy Model \cite{PhysRevB.24.2613,bouchaud1997universality} or the Parabolic Anderson Model \cite{van2006universality,gartner2005parabolic}), distributions with such exponential decay lay at the boundary between two different universality classes, and we surmise that the disappearance of the lowest bound state might have a deeper, statistical, meaning. Enlarging the present derivation to disorders with stretched exponential or even power-law tails, would however require a different approach.

\section{Conclusion}

In the present work, we have developed a mean-field approach to growth models with temporally correlated disorder. We extended the scope of the well-known travelling wave equation approach, building on work done in \cite{gueudre2014explore}. This method allows for a detailed analysis for a general It\={o} processes, and even leads to exact formulas of growth rates for a variety of disorders. We gave three examples, with gaussian, uniform or Laplace stationary distributions. It unveils universal features in growth from microscopic details, in particular in the small and large diffusion regimes. This suggests a methodology to fit such models on experimental data. The mean-field computation presents both an optimal growth point and a distinct freezing transition, features that have been also observed in many finite dimension models. In the present case, the optimal growth always lays in the quenched phase but a more detailed study of the statistics of $Z_i$ is dearly needed, and should be possible along the lines of \cite{DerridaSpohn1988}.

To match the numerous directions more phenomenological approaches of growth have taken, we suggest possible extensions of the present study. We wonder how to extend the analysis to heavy-tailed disorders, as they are now recognized as crucial ingredients of the large sensitivity of growth to environmental, financial or economic shocks \cite{sornette2006critical}. On the same side, the effect of non-stationary environments, adding a temporal dependence to the It\={o} equation itself, would further our understanding of delayed effects also commonly observed, such as population “momentum” \cite{lutz2003europe}. 

Finally we return to the primary motivations of the "polymers on tree", a spin glass toy model, and surmise our analysis could be made as rigorous as the original, white noise case \cite{arguin2013extremal}, an important step towards a theory of such processes. Nonetheless, those models are often treated with the replica tool, a very different and general approach, up to now limited to white-noise disorder. A better understanding of the above derivation in the language of replicas might open many other disordered systems to colored disorder.

\textbf{Acknowledgements.} We thank Alexander Dobrinevsky and Jean-Philippe Bouchaud for starting this line of thoughs, and many stimulating discussions. 

\bibliography{itopoly}

\begin{thebibliography}{54}%
\makeatletter
\providecommand \@ifxundefined [1]{%
 \@ifx{#1\undefined}
}%
\providecommand \@ifnum [1]{%
 \ifnum #1\expandafter \@firstoftwo
 \else \expandafter \@secondoftwo
 \fi
}%
\providecommand \@ifx [1]{%
 \ifx #1\expandafter \@firstoftwo
 \else \expandafter \@secondoftwo
 \fi
}%
\providecommand \natexlab [1]{#1}%
\providecommand \enquote  [1]{``#1''}%
\providecommand \bibnamefont  [1]{#1}%
\providecommand \bibfnamefont [1]{#1}%
\providecommand \citenamefont [1]{#1}%
\providecommand \href@noop [0]{\@secondoftwo}%
\providecommand \href [0]{\begingroup \@sanitize@url \@href}%
\providecommand \@href[1]{\@@startlink{#1}\@@href}%
\providecommand \@@href[1]{\endgroup#1\@@endlink}%
\providecommand \@sanitize@url [0]{\catcode `\\12\catcode `\$12\catcode
  `\&12\catcode `\#12\catcode `\^12\catcode `\_12\catcode `\%12\relax}%
\providecommand \@@startlink[1]{}%
\providecommand \@@endlink[0]{}%
\providecommand \url  [0]{\begingroup\@sanitize@url \@url }%
\providecommand \@url [1]{\endgroup\@href {#1}{\urlprefix }}%
\providecommand \urlprefix  [0]{URL }%
\providecommand \Eprint [0]{\href }%
\providecommand \doibase [0]{http://dx.doi.org/}%
\providecommand \selectlanguage [0]{\@gobble}%
\providecommand \bibinfo  [0]{\@secondoftwo}%
\providecommand \bibfield  [0]{\@secondoftwo}%
\providecommand \translation [1]{[#1]}%
\providecommand \BibitemOpen [0]{}%
\providecommand \bibitemStop [0]{}%
\providecommand \bibitemNoStop [0]{.\EOS\space}%
\providecommand \EOS [0]{\spacefactor3000\relax}%
\providecommand \BibitemShut  [1]{\csname bibitem#1\endcsname}%
\let\auto@bib@innerbib\@empty
\bibitem [{\citenamefont {Grossman}\ and\ \citenamefont
  {Helpman}(1993)}]{grossman1993innovation}%
  \BibitemOpen
  \bibfield  {author} {\bibinfo {author} {\bibfnamefont {G.~M.}\ \bibnamefont
  {Grossman}}\ and\ \bibinfo {author} {\bibfnamefont {E.}~\bibnamefont
  {Helpman}},\ }\href@noop {} {\emph {\bibinfo {title} {Innovation and growth
  in the global economy}}}\ (\bibinfo  {publisher} {MIT press},\ \bibinfo
  {year} {1993})\BibitemShut {NoStop}%
\bibitem [{\citenamefont {Cavalcanti}\ \emph {et~al.}(2015)\citenamefont
  {Cavalcanti}, \citenamefont {Tiago}, \citenamefont {Mohaddes},\ and\
  \citenamefont {Raissi}}]{cavalcanti2015commodity}%
  \BibitemOpen
  \bibfield  {author} {\bibinfo {author} {\bibfnamefont {D.~V.}\ \bibnamefont
  {Cavalcanti}}, \bibinfo {author} {\bibfnamefont {V.}~\bibnamefont {Tiago}},
  \bibinfo {author} {\bibfnamefont {K.}~\bibnamefont {Mohaddes}}, \ and\
  \bibinfo {author} {\bibfnamefont {M.}~\bibnamefont {Raissi}},\ }\href@noop {}
  {\bibfield  {journal} {\bibinfo  {journal} {Journal of Applied Econometrics}\
  }\textbf {\bibinfo {volume} {30}},\ \bibinfo {pages} {857} (\bibinfo {year}
  {2015})}\BibitemShut {NoStop}%
\bibitem [{\citenamefont {Bouchaud}(2015)}]{bouchaud2015growth}%
  \BibitemOpen
  \bibfield  {author} {\bibinfo {author} {\bibfnamefont {J.-P.}\ \bibnamefont
  {Bouchaud}},\ }\href@noop {} {\bibfield  {journal} {\bibinfo  {journal}
  {Journal of Statistical Mechanics: Theory and Experiment}\ }\textbf {\bibinfo
  {volume} {2015}},\ \bibinfo {pages} {P11011} (\bibinfo {year}
  {2015})}\BibitemShut {NoStop}%
\bibitem [{\citenamefont {Fat{\'a}s}\ and\ \citenamefont
  {Mihov}(2013)}]{fatas2013policy}%
  \BibitemOpen
  \bibfield  {author} {\bibinfo {author} {\bibfnamefont {A.}~\bibnamefont
  {Fat{\'a}s}}\ and\ \bibinfo {author} {\bibfnamefont {I.}~\bibnamefont
  {Mihov}},\ }\href@noop {} {\bibfield  {journal} {\bibinfo  {journal} {Review
  of Economics and Statistics}\ }\textbf {\bibinfo {volume} {95}},\ \bibinfo
  {pages} {362} (\bibinfo {year} {2013})}\BibitemShut {NoStop}%
\bibitem [{\citenamefont {Monod}(1949)}]{monod1949growth}%
  \BibitemOpen
  \bibfield  {author} {\bibinfo {author} {\bibfnamefont {J.}~\bibnamefont
  {Monod}},\ }\href@noop {} {\bibfield  {journal} {\bibinfo  {journal} {Annual
  Reviews in Microbiology}\ }\textbf {\bibinfo {volume} {3}},\ \bibinfo {pages}
  {371} (\bibinfo {year} {1949})}\BibitemShut {NoStop}%
\bibitem [{\citenamefont {Brunet}\ and\ \citenamefont
  {Derrida}(2012)}]{brunet2012genealogies}%
  \BibitemOpen
  \bibfield  {author} {\bibinfo {author} {\bibfnamefont {{\'E}.}~\bibnamefont
  {Brunet}}\ and\ \bibinfo {author} {\bibfnamefont {B.}~\bibnamefont
  {Derrida}},\ }\href@noop {} {\bibfield  {journal} {\bibinfo  {journal}
  {Philosophical Magazine}\ }\textbf {\bibinfo {volume} {92}},\ \bibinfo
  {pages} {255} (\bibinfo {year} {2012})}\BibitemShut {NoStop}%
\bibitem [{\citenamefont {Taleb}(2007)}]{taleb2007black}%
  \BibitemOpen
  \bibfield  {author} {\bibinfo {author} {\bibfnamefont {N.~N.}\ \bibnamefont
  {Taleb}},\ }\href@noop {} {\emph {\bibinfo {title} {The black swan: The
  impact of the highly improbable}}}\ (\bibinfo  {publisher} {Random house},\
  \bibinfo {year} {2007})\BibitemShut {NoStop}%
\bibitem [{\citenamefont {Chupeau}\ \emph {et~al.}(2016)\citenamefont
  {Chupeau}, \citenamefont {Benichou},\ and\ \citenamefont
  {Redner}}]{chupeau2016random}%
  \BibitemOpen
  \bibfield  {author} {\bibinfo {author} {\bibfnamefont {M.}~\bibnamefont
  {Chupeau}}, \bibinfo {author} {\bibfnamefont {O.}~\bibnamefont {Benichou}}, \
  and\ \bibinfo {author} {\bibfnamefont {S.}~\bibnamefont {Redner}},\
  }\href@noop {} {\bibfield  {journal} {\bibinfo  {journal} {arXiv preprint
  arXiv:1609.05151}\ } (\bibinfo {year} {2016})}\BibitemShut {NoStop}%
\bibitem [{\citenamefont {Bouchaud}\ and\ \citenamefont
  {Potters}(2003)}]{bouchaud2003theory}%
  \BibitemOpen
  \bibfield  {author} {\bibinfo {author} {\bibfnamefont {J.-P.}\ \bibnamefont
  {Bouchaud}}\ and\ \bibinfo {author} {\bibfnamefont {M.}~\bibnamefont
  {Potters}},\ }\href@noop {} {\emph {\bibinfo {title} {Theory of financial
  risk and derivative pricing: from statistical physics to risk management}}}\
  (\bibinfo  {publisher} {Cambridge university press},\ \bibinfo {year}
  {2003})\BibitemShut {NoStop}%
\bibitem [{\citenamefont {Lai}(2015)}]{lai2015giraffe}%
  \BibitemOpen
  \bibfield  {author} {\bibinfo {author} {\bibfnamefont {M.}~\bibnamefont
  {Lai}},\ }\href@noop {} {\bibfield  {journal} {\bibinfo  {journal} {arXiv
  preprint arXiv:1509.01549}\ } (\bibinfo {year} {2015})}\BibitemShut {NoStop}%
\bibitem [{\citenamefont {Cohen}\ \emph {et~al.}(2007)\citenamefont {Cohen},
  \citenamefont {McClure},\ and\ \citenamefont {Angela}}]{cohen2007should}%
  \BibitemOpen
  \bibfield  {author} {\bibinfo {author} {\bibfnamefont {J.~D.}\ \bibnamefont
  {Cohen}}, \bibinfo {author} {\bibfnamefont {S.~M.}\ \bibnamefont {McClure}},
  \ and\ \bibinfo {author} {\bibfnamefont {J.~Y.}\ \bibnamefont {Angela}},\
  }\href@noop {} {\bibfield  {journal} {\bibinfo  {journal} {Philosophical
  Transactions of the Royal Society of London B: Biological Sciences}\ }\textbf
  {\bibinfo {volume} {362}},\ \bibinfo {pages} {933} (\bibinfo {year}
  {2007})}\BibitemShut {NoStop}%
\bibitem [{\citenamefont {Vespignani}(2012)}]{vespignani2012modelling}%
  \BibitemOpen
  \bibfield  {author} {\bibinfo {author} {\bibfnamefont {A.}~\bibnamefont
  {Vespignani}},\ }\href@noop {} {\bibfield  {journal} {\bibinfo  {journal}
  {Nature Physics}\ }\textbf {\bibinfo {volume} {8}},\ \bibinfo {pages} {32}
  (\bibinfo {year} {2012})}\BibitemShut {NoStop}%
\bibitem [{\citenamefont {Sutton}\ and\ \citenamefont
  {Barto}(1998)}]{sutton1998reinforcement}%
  \BibitemOpen
  \bibfield  {author} {\bibinfo {author} {\bibfnamefont {R.~S.}\ \bibnamefont
  {Sutton}}\ and\ \bibinfo {author} {\bibfnamefont {A.~G.}\ \bibnamefont
  {Barto}},\ }\href@noop {} {\emph {\bibinfo {title} {Reinforcement learning:
  An introduction}}},\ Vol.~\bibinfo {volume} {1}\ (\bibinfo  {publisher} {MIT
  press Cambridge},\ \bibinfo {year} {1998})\BibitemShut {NoStop}%
\bibitem [{\citenamefont {Pratt}\ and\ \citenamefont
  {Sumpter}(2006)}]{pratt2006tunable}%
  \BibitemOpen
  \bibfield  {author} {\bibinfo {author} {\bibfnamefont {S.~C.}\ \bibnamefont
  {Pratt}}\ and\ \bibinfo {author} {\bibfnamefont {D.~J.}\ \bibnamefont
  {Sumpter}},\ }\href@noop {} {\bibfield  {journal} {\bibinfo  {journal}
  {Proceedings of the National Academy of Sciences}\ }\textbf {\bibinfo
  {volume} {103}},\ \bibinfo {pages} {15906} (\bibinfo {year}
  {2006})}\BibitemShut {NoStop}%
\bibitem [{\citenamefont {Tokic}(2010)}]{tokic2010adaptive}%
  \BibitemOpen
  \bibfield  {author} {\bibinfo {author} {\bibfnamefont {M.}~\bibnamefont
  {Tokic}},\ }in\ \href@noop {} {\emph {\bibinfo {booktitle} {Annual Conference
  on Artificial Intelligence}}}\ (\bibinfo {organization} {Springer},\ \bibinfo
  {year} {2010})\ pp.\ \bibinfo {pages} {203--210}\BibitemShut {NoStop}%
\bibitem [{\citenamefont {Gueudr{\'e}}\ \emph {et~al.}(2014)\citenamefont
  {Gueudr{\'e}}, \citenamefont {Dobrinevski},\ and\ \citenamefont
  {Bouchaud}}]{gueudre2014explore}%
  \BibitemOpen
  \bibfield  {author} {\bibinfo {author} {\bibfnamefont {T.}~\bibnamefont
  {Gueudr{\'e}}}, \bibinfo {author} {\bibfnamefont {A.}~\bibnamefont
  {Dobrinevski}}, \ and\ \bibinfo {author} {\bibfnamefont {J.-P.}\ \bibnamefont
  {Bouchaud}},\ }\href@noop {} {\bibfield  {journal} {\bibinfo  {journal}
  {Physical review letters}\ }\textbf {\bibinfo {volume} {112}},\ \bibinfo
  {pages} {050602} (\bibinfo {year} {2014})}\BibitemShut {NoStop}%
\bibitem [{\citenamefont {Derrida}\ and\ \citenamefont
  {Spohn}(1988)}]{DerridaSpohn1988}%
  \BibitemOpen
  \bibfield  {author} {\bibinfo {author} {\bibfnamefont {B.}~\bibnamefont
  {Derrida}}\ and\ \bibinfo {author} {\bibfnamefont {H.}~\bibnamefont
  {Spohn}},\ }\href {\doibase 10.1007/BF01014886} {\bibfield  {journal}
  {\bibinfo  {journal} {J Stat Phys}\ }\textbf {\bibinfo {volume} {51}},\
  \bibinfo {pages} {817} (\bibinfo {year} {1988})}\BibitemShut {NoStop}%
\bibitem [{\citenamefont {Brunet}\ and\ \citenamefont
  {Derrida}(2011)}]{brunet2011branching}%
  \BibitemOpen
  \bibfield  {author} {\bibinfo {author} {\bibfnamefont {{\'E}.}~\bibnamefont
  {Brunet}}\ and\ \bibinfo {author} {\bibfnamefont {B.}~\bibnamefont
  {Derrida}},\ }\href@noop {} {\bibfield  {journal} {\bibinfo  {journal}
  {Journal of Statistical Physics}\ }\textbf {\bibinfo {volume} {143}},\
  \bibinfo {pages} {420} (\bibinfo {year} {2011})}\BibitemShut {NoStop}%
\bibitem [{\citenamefont {Grindrod}(1996)}]{grindrod1996theory}%
  \BibitemOpen
  \bibfield  {author} {\bibinfo {author} {\bibfnamefont {P.}~\bibnamefont
  {Grindrod}},\ }\href@noop {} {\emph {\bibinfo {title} {The theory and
  applications of reaction-diffusion equations: patterns and waves}}}\
  (\bibinfo  {publisher} {Clarendon Press},\ \bibinfo {year}
  {1996})\BibitemShut {NoStop}%
\bibitem [{\citenamefont {Arguin}\ \emph {et~al.}(2013)\citenamefont {Arguin},
  \citenamefont {Bovier},\ and\ \citenamefont {Kistler}}]{arguin2013extremal}%
  \BibitemOpen
  \bibfield  {author} {\bibinfo {author} {\bibfnamefont {L.-P.}\ \bibnamefont
  {Arguin}}, \bibinfo {author} {\bibfnamefont {A.}~\bibnamefont {Bovier}}, \
  and\ \bibinfo {author} {\bibfnamefont {N.}~\bibnamefont {Kistler}},\
  }\href@noop {} {\bibfield  {journal} {\bibinfo  {journal} {Probability Theory
  and related fields}\ }\textbf {\bibinfo {volume} {157}},\ \bibinfo {pages}
  {535} (\bibinfo {year} {2013})}\BibitemShut {NoStop}%
\bibitem [{\citenamefont {Gittins}\ and\ \citenamefont {Jones}(1974)}]{Exp3}%
  \BibitemOpen
  \bibfield  {author} {\bibinfo {author} {\bibfnamefont {J.~C.}\ \bibnamefont
  {Gittins}}\ and\ \bibinfo {author} {\bibfnamefont {D.~M.}\ \bibnamefont
  {Jones}},\ }in\ \href {http://www.ams.org/mathscinet-getitem?mr=0370964}
  {\emph {\bibinfo {booktitle} {Progress in statistics (European Meeting
  Statisticians, Budapest, 1972)}}}\ (\bibinfo  {publisher} {North-Holland},\
  \bibinfo {address} {Amsterdam},\ \bibinfo {year} {1974})\ pp.\ \bibinfo
  {pages} {241--266}\BibitemShut {NoStop}%
\bibitem [{\citenamefont {Gardiner}\ \emph {et~al.}(1985)\citenamefont
  {Gardiner} \emph {et~al.}}]{gardiner1985handbook}%
  \BibitemOpen
  \bibfield  {author} {\bibinfo {author} {\bibfnamefont {C.~W.}\ \bibnamefont
  {Gardiner}} \emph {et~al.},\ }\href@noop {} {\emph {\bibinfo {title}
  {Handbook of stochastic methods}}},\ Vol.~\bibinfo {volume} {3}\ (\bibinfo
  {publisher} {Springer Berlin},\ \bibinfo {year} {1985})\BibitemShut {NoStop}%
\bibitem [{\citenamefont {Pavliotis}(2014)}]{pavliotis2014stochastic}%
  \BibitemOpen
  \bibfield  {author} {\bibinfo {author} {\bibfnamefont {G.~A.}\ \bibnamefont
  {Pavliotis}},\ }\href@noop {} {\bibfield  {journal} {\bibinfo  {journal}
  {Diffusion Processes, the Fokker-Planck}\ } (\bibinfo {year}
  {2014})}\BibitemShut {NoStop}%
\bibitem [{\citenamefont {Gomme}(1993)}]{gomme1993money}%
  \BibitemOpen
  \bibfield  {author} {\bibinfo {author} {\bibfnamefont {P.}~\bibnamefont
  {Gomme}},\ }\href@noop {} {\bibfield  {journal} {\bibinfo  {journal} {Journal
  of Monetary economics}\ }\textbf {\bibinfo {volume} {32}},\ \bibinfo {pages}
  {51} (\bibinfo {year} {1993})}\BibitemShut {NoStop}%
\bibitem [{\citenamefont {Cooley}\ and\ \citenamefont
  {Prescott}(1995)}]{cooley1995economic}%
  \BibitemOpen
  \bibfield  {author} {\bibinfo {author} {\bibfnamefont {T.~F.}\ \bibnamefont
  {Cooley}}\ and\ \bibinfo {author} {\bibfnamefont {E.~C.}\ \bibnamefont
  {Prescott}},\ }\href@noop {} {\bibfield  {journal} {\bibinfo  {journal}
  {Frontiers of business cycle research}\ }\textbf {\bibinfo {volume} {1}}
  (\bibinfo {year} {1995})}\BibitemShut {NoStop}%
\bibitem [{\citenamefont {Janssen}\ and\ \citenamefont
  {Heuberger}(1995)}]{janssen1995calibration}%
  \BibitemOpen
  \bibfield  {author} {\bibinfo {author} {\bibfnamefont {P.}~\bibnamefont
  {Janssen}}\ and\ \bibinfo {author} {\bibfnamefont {P.}~\bibnamefont
  {Heuberger}},\ }\href@noop {} {\bibfield  {journal} {\bibinfo  {journal}
  {Ecological Modelling}\ }\textbf {\bibinfo {volume} {83}},\ \bibinfo {pages}
  {55} (\bibinfo {year} {1995})}\BibitemShut {NoStop}%
\bibitem [{\citenamefont {Caraglio}\ \emph {et~al.}(2016)\citenamefont
  {Caraglio}, \citenamefont {Baldovin},\ and\ \citenamefont
  {Stella}}]{caraglio2016export}%
  \BibitemOpen
  \bibfield  {author} {\bibinfo {author} {\bibfnamefont {M.}~\bibnamefont
  {Caraglio}}, \bibinfo {author} {\bibfnamefont {F.}~\bibnamefont {Baldovin}},
  \ and\ \bibinfo {author} {\bibfnamefont {A.~L.}\ \bibnamefont {Stella}},\
  }\href@noop {} {\bibfield  {journal} {\bibinfo  {journal} {Scientific
  Reports}\ }\textbf {\bibinfo {volume} {6}} (\bibinfo {year}
  {2016})}\BibitemShut {NoStop}%
\bibitem [{\citenamefont {Fyodorov}\ and\ \citenamefont
  {Bouchaud}(2008)}]{fyodorov2008freezing}%
  \BibitemOpen
  \bibfield  {author} {\bibinfo {author} {\bibfnamefont {Y.~V.}\ \bibnamefont
  {Fyodorov}}\ and\ \bibinfo {author} {\bibfnamefont {J.-P.}\ \bibnamefont
  {Bouchaud}},\ }\href@noop {} {\bibfield  {journal} {\bibinfo  {journal}
  {Journal of Physics A: Mathematical and Theoretical}\ }\textbf {\bibinfo
  {volume} {41}},\ \bibinfo {pages} {372001} (\bibinfo {year}
  {2008})}\BibitemShut {NoStop}%
\bibitem [{\citenamefont {Fyodorov}\ \emph {et~al.}(2009)\citenamefont
  {Fyodorov}, \citenamefont {Le~Doussal},\ and\ \citenamefont
  {Rosso}}]{fyodorov2009statistical}%
  \BibitemOpen
  \bibfield  {author} {\bibinfo {author} {\bibfnamefont {Y.~V.}\ \bibnamefont
  {Fyodorov}}, \bibinfo {author} {\bibfnamefont {P.}~\bibnamefont
  {Le~Doussal}}, \ and\ \bibinfo {author} {\bibfnamefont {A.}~\bibnamefont
  {Rosso}},\ }\href@noop {} {\bibfield  {journal} {\bibinfo  {journal} {Journal
  of Statistical Mechanics: Theory and Experiment}\ }\textbf {\bibinfo {volume}
  {2009}},\ \bibinfo {pages} {P10005} (\bibinfo {year} {2009})}\BibitemShut
  {NoStop}%
\bibitem [{\citenamefont {Zamolodchikov}\ and\ \citenamefont
  {Zamolodchikov}(2007)}]{zamolodchikov2007lectures}%
  \BibitemOpen
  \bibfield  {author} {\bibinfo {author} {\bibfnamefont {A.}~\bibnamefont
  {Zamolodchikov}}\ and\ \bibinfo {author} {\bibfnamefont {A.}~\bibnamefont
  {Zamolodchikov}},\ }\href@noop {} {\enquote {\bibinfo {title} {Lectures on
  liouville theory and matrix models},}\ } (\bibinfo {year} {2007})\BibitemShut
  {NoStop}%
\bibitem [{\citenamefont {Fyodorov}\ and\ \citenamefont
  {Keating}(2014)}]{fyodorov2014freezing}%
  \BibitemOpen
  \bibfield  {author} {\bibinfo {author} {\bibfnamefont {Y.~V.}\ \bibnamefont
  {Fyodorov}}\ and\ \bibinfo {author} {\bibfnamefont {J.~P.}\ \bibnamefont
  {Keating}},\ }\href@noop {} {\bibfield  {journal} {\bibinfo  {journal}
  {Philosophical Transactions of the Royal Society of London A: Mathematical,
  Physical and Engineering Sciences}\ }\textbf {\bibinfo {volume} {372}},\
  \bibinfo {pages} {20120503} (\bibinfo {year} {2014})}\BibitemShut {NoStop}%
\bibitem [{\citenamefont {Cooper}\ \emph {et~al.}(1995)\citenamefont {Cooper},
  \citenamefont {Khare},\ and\ \citenamefont
  {Sukhatme}}]{cooper1995supersymmetry}%
  \BibitemOpen
  \bibfield  {author} {\bibinfo {author} {\bibfnamefont {F.}~\bibnamefont
  {Cooper}}, \bibinfo {author} {\bibfnamefont {A.}~\bibnamefont {Khare}}, \
  and\ \bibinfo {author} {\bibfnamefont {U.}~\bibnamefont {Sukhatme}},\
  }\href@noop {} {\bibfield  {journal} {\bibinfo  {journal} {Physics Reports}\
  }\textbf {\bibinfo {volume} {251}},\ \bibinfo {pages} {267} (\bibinfo {year}
  {1995})}\BibitemShut {NoStop}%
\bibitem [{\citenamefont {Risken}(1984)}]{risken1984fokker}%
  \BibitemOpen
  \bibfield  {author} {\bibinfo {author} {\bibfnamefont {H.}~\bibnamefont
  {Risken}},\ }in\ \href@noop {} {\emph {\bibinfo {booktitle} {The
  Fokker-Planck Equation}}}\ (\bibinfo  {publisher} {Springer},\ \bibinfo
  {year} {1984})\ pp.\ \bibinfo {pages} {63--95}\BibitemShut {NoStop}%
\bibitem [{\citenamefont {Jung}\ and\ \citenamefont
  {Risken}(1985)}]{jung1985correlation}%
  \BibitemOpen
  \bibfield  {author} {\bibinfo {author} {\bibfnamefont {P.}~\bibnamefont
  {Jung}}\ and\ \bibinfo {author} {\bibfnamefont {H.}~\bibnamefont {Risken}},\
  }\href@noop {} {\bibfield  {journal} {\bibinfo  {journal} {Zeitschrift
  f{\"u}r Physik B Condensed Matter}\ }\textbf {\bibinfo {volume} {59}},\
  \bibinfo {pages} {469} (\bibinfo {year} {1985})}\BibitemShut {NoStop}%
\bibitem [{\citenamefont {Cook}\ and\ \citenamefont
  {Derrida}(1989)}]{CookDerrida1990}%
  \BibitemOpen
  \bibfield  {author} {\bibinfo {author} {\bibfnamefont {J.}~\bibnamefont
  {Cook}}\ and\ \bibinfo {author} {\bibfnamefont {B.}~\bibnamefont {Derrida}},\
  }\href {\doibase 10.1007/BF01023636} {\bibfield  {journal} {\bibinfo
  {journal} {Journal of Statistical Physics}\ }\textbf {\bibinfo {volume}
  {57}},\ \bibinfo {pages} {89} (\bibinfo {year} {1989})}\BibitemShut {NoStop}%
\bibitem [{\citenamefont {Economou}(1984)}]{economou1984green}%
  \BibitemOpen
  \bibfield  {author} {\bibinfo {author} {\bibfnamefont {E.~N.}\ \bibnamefont
  {Economou}},\ }\href@noop {} {\emph {\bibinfo {title} {Green's functions in
  quantum physics}}},\ Vol.~\bibinfo {volume} {3}\ (\bibinfo  {publisher}
  {Springer},\ \bibinfo {year} {1984})\BibitemShut {NoStop}%
\bibitem [{\citenamefont {Bramson}(1983)}]{bramson1983convergence}%
  \BibitemOpen
  \bibfield  {author} {\bibinfo {author} {\bibfnamefont {M.}~\bibnamefont
  {Bramson}},\ }\href@noop {} {\emph {\bibinfo {title} {Convergence of
  solutions of the Kolmogorov equation to travelling waves}}},\ Vol.\ \bibinfo
  {volume} {285}\ (\bibinfo  {publisher} {American Mathematical Soc.},\
  \bibinfo {year} {1983})\BibitemShut {NoStop}%
\bibitem [{\citenamefont {Schiff}(1955)}]{schiff1955quantum}%
  \BibitemOpen
  \bibfield  {author} {\bibinfo {author} {\bibfnamefont {L.}~\bibnamefont
  {Schiff}},\ }\href {https://books.google.it/books?id=7ApRAAAAMAAJ} {\emph
  {\bibinfo {title} {Quantum Mechanics}}},\ International series in pure and
  applied physics\ (\bibinfo  {publisher} {McGraw-Hill},\ \bibinfo {year}
  {1955})\BibitemShut {NoStop}%
\bibitem [{\citenamefont {Yor}(2012)}]{yor2012exponential}%
  \BibitemOpen
  \bibfield  {author} {\bibinfo {author} {\bibfnamefont {M.}~\bibnamefont
  {Yor}},\ }\href@noop {} {\emph {\bibinfo {title} {Exponential functionals of
  Brownian motion and related processes}}}\ (\bibinfo  {publisher} {Springer
  Science \& Business Media},\ \bibinfo {year} {2012})\BibitemShut {NoStop}%
\bibitem [{\citenamefont {Rogers}\ and\ \citenamefont
  {Williams}(2000)}]{rogers2000diffusions}%
  \BibitemOpen
  \bibfield  {author} {\bibinfo {author} {\bibfnamefont {L.~C.~G.}\
  \bibnamefont {Rogers}}\ and\ \bibinfo {author} {\bibfnamefont
  {D.}~\bibnamefont {Williams}},\ }\href@noop {} {\emph {\bibinfo {title}
  {Diffusions, Markov processes and martingales: Volume 2, It{\^o}
  calculus}}},\ Vol.~\bibinfo {volume} {2}\ (\bibinfo  {publisher} {Cambridge
  university press},\ \bibinfo {year} {2000})\BibitemShut {NoStop}%
\bibitem [{\citenamefont {Robinett}(2009)}]{robinett2009stark}%
  \BibitemOpen
  \bibfield  {author} {\bibinfo {author} {\bibfnamefont {R.}~\bibnamefont
  {Robinett}},\ }\href@noop {} {\bibfield  {journal} {\bibinfo  {journal}
  {European Journal of Physics}\ }\textbf {\bibinfo {volume} {31}},\ \bibinfo
  {pages} {1} (\bibinfo {year} {2009})}\BibitemShut {NoStop}%
\bibitem [{\citenamefont {Angel}\ and\ \citenamefont
  {Sandars}(1968)}]{angel1968hyperfine}%
  \BibitemOpen
  \bibfield  {author} {\bibinfo {author} {\bibfnamefont {J.}~\bibnamefont
  {Angel}}\ and\ \bibinfo {author} {\bibfnamefont {P.}~\bibnamefont
  {Sandars}},\ }in\ \href@noop {} {\emph {\bibinfo {booktitle} {Proceedings of
  the Royal Society of London A: Mathematical, Physical and Engineering
  Sciences}}},\ Vol.\ \bibinfo {volume} {305}\ (\bibinfo {organization} {The
  Royal Society},\ \bibinfo {year} {1968})\ pp.\ \bibinfo {pages}
  {125--138}\BibitemShut {NoStop}%
\bibitem [{\citenamefont {Bastard}\ \emph {et~al.}(1983)\citenamefont
  {Bastard}, \citenamefont {Mendez}, \citenamefont {Chang},\ and\ \citenamefont
  {Esaki}}]{bastard1983variational}%
  \BibitemOpen
  \bibfield  {author} {\bibinfo {author} {\bibfnamefont {G.}~\bibnamefont
  {Bastard}}, \bibinfo {author} {\bibfnamefont {E.}~\bibnamefont {Mendez}},
  \bibinfo {author} {\bibfnamefont {L.}~\bibnamefont {Chang}}, \ and\ \bibinfo
  {author} {\bibfnamefont {L.}~\bibnamefont {Esaki}},\ }\href@noop {}
  {\bibfield  {journal} {\bibinfo  {journal} {Physical Review B}\ }\textbf
  {\bibinfo {volume} {28}},\ \bibinfo {pages} {3241} (\bibinfo {year}
  {1983})}\BibitemShut {NoStop}%
\bibitem [{\citenamefont {Merzbacher}(1970)}]{merzbacher1970quantum}%
  \BibitemOpen
  \bibfield  {author} {\bibinfo {author} {\bibfnamefont {E.}~\bibnamefont
  {Merzbacher}},\ }\href@noop {} {\enquote {\bibinfo {title} {Quantum
  mechanics},}\ } (\bibinfo {year} {1970})\BibitemShut {NoStop}%
\bibitem [{\citenamefont {Bender}\ and\ \citenamefont
  {Orszag}(1999)}]{bender1999advanced}%
  \BibitemOpen
  \bibfield  {author} {\bibinfo {author} {\bibfnamefont {C.~M.}\ \bibnamefont
  {Bender}}\ and\ \bibinfo {author} {\bibfnamefont {S.~A.}\ \bibnamefont
  {Orszag}},\ }\href@noop {} {\emph {\bibinfo {title} {Advanced mathematical
  methods for scientists and engineers I}}}\ (\bibinfo  {publisher} {Springer
  Science \& Business Media},\ \bibinfo {year} {1999})\BibitemShut {NoStop}%
\bibitem [{\citenamefont {Sornette}(2006)}]{sornette2006critical}%
  \BibitemOpen
  \bibfield  {author} {\bibinfo {author} {\bibfnamefont {D.}~\bibnamefont
  {Sornette}},\ }\href@noop {} {\emph {\bibinfo {title} {Critical phenomena in
  natural sciences: chaos, fractals, selforganization and disorder: concepts
  and tools}}}\ (\bibinfo  {publisher} {Springer Science \& Business Media},\
  \bibinfo {year} {2006})\BibitemShut {NoStop}%
\bibitem [{\citenamefont {Newman}(1996)}]{newman1996self}%
  \BibitemOpen
  \bibfield  {author} {\bibinfo {author} {\bibfnamefont {M.}~\bibnamefont
  {Newman}},\ }\href@noop {} {\bibfield  {journal} {\bibinfo  {journal}
  {Proceedings of the Royal Society of London B: Biological Sciences}\ }\textbf
  {\bibinfo {volume} {263}},\ \bibinfo {pages} {1605} (\bibinfo {year}
  {1996})}\BibitemShut {NoStop}%
\bibitem [{\citenamefont {Mora}\ and\ \citenamefont
  {Bialek}(2011)}]{mora2011biological}%
  \BibitemOpen
  \bibfield  {author} {\bibinfo {author} {\bibfnamefont {T.}~\bibnamefont
  {Mora}}\ and\ \bibinfo {author} {\bibfnamefont {W.}~\bibnamefont {Bialek}},\
  }\href@noop {} {\bibfield  {journal} {\bibinfo  {journal} {Journal of
  Statistical Physics}\ }\textbf {\bibinfo {volume} {144}},\ \bibinfo {pages}
  {268} (\bibinfo {year} {2011})}\BibitemShut {NoStop}%
\bibitem [{\citenamefont {Lukes}\ and\ \citenamefont
  {Somaratna}(1969)}]{lukes1969exact}%
  \BibitemOpen
  \bibfield  {author} {\bibinfo {author} {\bibfnamefont {T.}~\bibnamefont
  {Lukes}}\ and\ \bibinfo {author} {\bibfnamefont {K.}~\bibnamefont
  {Somaratna}},\ }\href@noop {} {\bibfield  {journal} {\bibinfo  {journal}
  {Journal of Physics C: Solid State Physics}\ }\textbf {\bibinfo {volume}
  {2}},\ \bibinfo {pages} {586} (\bibinfo {year} {1969})}\BibitemShut {NoStop}%
\bibitem [{\citenamefont {Derrida}(1981)}]{PhysRevB.24.2613}%
  \BibitemOpen
  \bibfield  {author} {\bibinfo {author} {\bibfnamefont {B.}~\bibnamefont
  {Derrida}},\ }\href {\doibase 10.1103/PhysRevB.24.2613} {\bibfield  {journal}
  {\bibinfo  {journal} {Phys. Rev. B}\ }\textbf {\bibinfo {volume} {24}},\
  \bibinfo {pages} {2613} (\bibinfo {year} {1981})}\BibitemShut {NoStop}%
\bibitem [{\citenamefont {Bouchaud}\ and\ \citenamefont
  {M{\'e}zard}(1997)}]{bouchaud1997universality}%
  \BibitemOpen
  \bibfield  {author} {\bibinfo {author} {\bibfnamefont {J.-P.}\ \bibnamefont
  {Bouchaud}}\ and\ \bibinfo {author} {\bibfnamefont {M.}~\bibnamefont
  {M{\'e}zard}},\ }\href@noop {} {\bibfield  {journal} {\bibinfo  {journal}
  {Journal of Physics A: Mathematical and General}\ }\textbf {\bibinfo {volume}
  {30}},\ \bibinfo {pages} {7997} (\bibinfo {year} {1997})}\BibitemShut
  {NoStop}%
\bibitem [{\citenamefont {Van Der~Hofstad}\ \emph {et~al.}(2006)\citenamefont
  {Van Der~Hofstad}, \citenamefont {K{\"o}nig},\ and\ \citenamefont
  {M{\"o}rters}}]{van2006universality}%
  \BibitemOpen
  \bibfield  {author} {\bibinfo {author} {\bibfnamefont {R.}~\bibnamefont {Van
  Der~Hofstad}}, \bibinfo {author} {\bibfnamefont {W.}~\bibnamefont
  {K{\"o}nig}}, \ and\ \bibinfo {author} {\bibfnamefont {P.}~\bibnamefont
  {M{\"o}rters}},\ }\href@noop {} {\bibfield  {journal} {\bibinfo  {journal}
  {Communications in mathematical physics}\ }\textbf {\bibinfo {volume}
  {267}},\ \bibinfo {pages} {307} (\bibinfo {year} {2006})}\BibitemShut
  {NoStop}%
\bibitem [{\citenamefont {G{\"a}rtner}\ and\ \citenamefont
  {K{\"o}nig}(2005)}]{gartner2005parabolic}%
  \BibitemOpen
  \bibfield  {author} {\bibinfo {author} {\bibfnamefont {J.}~\bibnamefont
  {G{\"a}rtner}}\ and\ \bibinfo {author} {\bibfnamefont {W.}~\bibnamefont
  {K{\"o}nig}},\ }in\ \href@noop {} {\emph {\bibinfo {booktitle} {Interacting
  stochastic systems}}}\ (\bibinfo  {publisher} {Springer},\ \bibinfo {year}
  {2005})\ pp.\ \bibinfo {pages} {153--179}\BibitemShut {NoStop}%
\bibitem [{\citenamefont {Lutz}\ \emph {et~al.}(2003)\citenamefont {Lutz},
  \citenamefont {O'Neill},\ and\ \citenamefont {Scherbov}}]{lutz2003europe}%
  \BibitemOpen
  \bibfield  {author} {\bibinfo {author} {\bibfnamefont {W.}~\bibnamefont
  {Lutz}}, \bibinfo {author} {\bibfnamefont {B.~C.}\ \bibnamefont {O'Neill}}, \
  and\ \bibinfo {author} {\bibfnamefont {S.}~\bibnamefont {Scherbov}},\
  }\href@noop {} {\bibfield  {journal} {\bibinfo  {journal} {Science}\ }\textbf
  {\bibinfo {volume} {299}},\ \bibinfo {pages} {1991} (\bibinfo {year}
  {2003})}\BibitemShut {NoStop}%
\end{thebibliography}%
\clearpage
\newpage

\appendix
\section{An alternative solution of the OU process \label{prl_OU}}

Another possibility \cite{gueudre2014explore} to solve The OU noise model is to write down the eigenvectors, and eigenvalues of the harmonic oscillator, and leave the implicit equation as a sum. 
The normalized eigenbasis is built over the Hermite functions, and given by:
\bea
\alpha_n &= k n-\frac{\gamma ^2 D_2}{k^2} \text{,   } n\geq 0 \nonumber \\
\phi_n &= \left( 2^n n! \sqrt{\frac{2 D_2\pi }{k}} \right)^{-1/2} e^{-\tilde{\eta}^2/2} H_n(\tilde{\eta}) \nonumber
\eea
with $H_n$ the Hermite polynomials. Remains to compute the projection of the eigenvectors over $Q_H$:
\bea
\langle Q_H | \phi_n \rangle ^2 = \frac{1}{n!} \left(\frac{ D_2 \gamma^2}{k^3}\right)^n e^{-D_2 \gamma^2 / k^3} \nonumber
\eea

Plugging this expression into Eq.\ref{final_res} finally leads to an implicit expression for the curve $c(\gamma)$:
\bea
\frac{2^{\gamma}}{\lambda} = e^{-D_2 \gamma^2 / k^3} \sum_{n=0} ^{\infty} \frac{\left( D_2 \gamma^2 / k^3 \right)^n}{n !(k n-\gamma ^2 D_2/k^2 - \hat{\lambda})}
\eea
It gives back the result from \cite{gueudre2014explore} with the convention $D_2 = \sigma ^2 / 2 \tau ^2$ and $k = 1/ \tau$.

\section{Monotonous decay of the annealed branch \label{monotonous}}

Here we show that the annealed branch $c_a(\lambda)$, according to Eq.\ref{final_res},  is necessarily a decreasing function of $\lambda$. Let us first recall Eq.\ref{final_res} in the annealed regime $\gamma =1$:
\begin{align}
\frac{2}{\lambda} =  \sum_{n \geq 0} \frac{\langle Q_H | \phi_n \rangle ^2}{\alpha_n +\lambda/2 + c(\lambda)}
\label{recall_final_res}
\end{align}
and derivate it w.r.t to $\lambda$:
\begin{align}
\frac{2}{\lambda^2} = \left(\frac{1}{2} + \frac{\partial c}{\partial \lambda} \right) \sum_{n \geq 0} \frac{\langle Q_H | \phi_n \rangle^2}{(\alpha_n +\lambda/2 + c(\lambda))^2} \nonumber
\end{align}
Substituting the left hand-side with Eq.\ref{recall_final_res}, we are left with:
\begin{align}
2  \frac{\partial c}{\partial \lambda} \sum_{n \geq 0} \frac{\langle Q_H | \phi_n \rangle^2}{(\alpha_n +\lambda/2 + c(\lambda))^2} &= \nonumber\\
\left(\sum_{n \geq 0} \frac{\langle Q_H | \phi_n \rangle^2}{\alpha_n +\lambda/2 + c(\lambda)} \right)^2 &- \sum_{n \geq 0} \frac{\langle Q_H | \phi_n \rangle^2}{(\alpha_n +\lambda/2 + c(\lambda))^2} \nonumber
\end{align}
But using the Cauchy-Schwart inequality over the first term of the right hand side:
\begin{align}
&\left(\sum_{n \geq 0} \frac{\langle Q_H | \phi_n \rangle^2}{\alpha_n +\lambda/2 + c(\lambda)} \right)^2 \nonumber \\
&= \left(\sum_{n \geq 0}\langle Q_H | \phi_n \rangle \times \frac{\langle Q_H | \phi_n \rangle}{\alpha_n +\lambda/2 + c(\lambda)} \right)^2 \nonumber\\
&\leq \left( \sum_n \langle Q_H | \phi_n \rangle ^2 \right) \left(\sum_n \frac{\langle Q_H | \phi_n \rangle^2}{(\alpha_n +\lambda/2 + c(\lambda))^2} \right)\nonumber \\
&\leq \sum_n \frac{\langle Q_H | \phi_n \rangle^2}{(\alpha_n +\lambda/2 + c(\lambda))^2}\nonumber
\end{align}
using the fact that:
\begin{align}
&\sum_n \langle Q_H | \phi_n \rangle ^2 = ||Q_H||_2 ^2 \nonumber \\
&=\frac{1}{\mathcal{N}} \int_\eta d\eta \exp(-\Phi(\eta))=1 \nonumber
\end{align}
and so:
\begin{align}
\frac{\partial c}{\partial \lambda}  \leq 0
\end{align}

\section{The exponential model \label{app_exp_model}}

The process with a Laplace stationary distribution represents a singular case in this class of models. It follows:
\begin{align}
V(\eta) &=  \frac{k^2}{4 D_2^3} - \frac{k}{D_2} \delta({\eta}) = A - B \delta({\eta}) \nonumber \\
f(\eta)&= k |x| \nonumber \\
Q(\eta) &= \frac{k}{2 D_2} \exp(-\frac{k}{D_2} |x|) \nonumber \\
Q_H(\eta) &=  \sqrt{\frac{k}{2 D_2}} \exp(-\frac{k}{2 D_2} |x|) \nonumber
\end{align}

For $\delta$ potentials, the Dyson equation can be solved exactly in coordinate representation, and gives the Green function $G_{\gamma}$ as a function of the well-known Green function, noted $G_0$, for the free particle under an electric field \cite{lukes1969exact}:
\begin{align}
G_{\gamma}(x,y;z) &= G_0(x,y;z) + \nonumber \\
\nonumber \\
&\frac{B \times G_0(x,0;z)G_0(0,y;z)}{1-B \times G_0(0,0;z)} \label{dyson_delta}\\
\nonumber\\
G_0(x,y;z) &= - i (2 \pi i)^{-1/2} \times \nonumber \\
\nonumber \\
\int_0 ^{\infty} t^{-1/2} &\exp \left[i \left(z t + \frac{(x+y)\gamma t}{2 D_2} - \frac{\gamma ^2 t^3}{24 D_2^2}\right) \right] dt \nonumber
\end{align}
Eq.\ref{dyson_delta} readily shows that no bound state survives to the electric field in a Dirac potential. Eq.\ref{final_res}, regularized by the addition of a small imaginary part $\hat{\lambda} \rightarrow \hat{\lambda} + i \epsilon$, reduces to (setting $D_2=1$ for simplicity):
\begin{align}
\int_0 ^{\infty} dt \frac{8 e^{i \hat{\lambda} \sqrt{t}} k^3 }{(k^2+t \gamma^2)^2} \left( e^{i t^{5/2} \gamma^2/24} - e^{i \hat{\lambda} \sqrt{t}}k\right)^{-1} = \frac{2^{\gamma}}{\lambda} \nonumber
\end{align}

Although the above equation should lead to the growth rate, the appearance of oscillating terms makes it unsuitable for numerical estimations, and we have been unable to confirm its validity.
\clearpage
\end{document}